\documentclass{emulateapj}
\usepackage{graphicx}
\usepackage{epsfig}
\usepackage{subfigure}
\slugcomment{Accepted for publication in ApJS COSMOS Special Issue, January 9, 2007}

\shorttitle{Morphology of star forming galaxies in COSMOS}
\shortauthors{Zamojski et al.}

\begin{document}

\title{Deep GALEX imaging of the HST/COSMOS field:  A first look at the morphology of \lowercase{$z \sim 0.7$} star-forming galaxies\altaffilmark{*}}

\author{
M. A. Zamojski\altaffilmark{1},
D. Schiminovich\altaffilmark{1},
R. M. Rich\altaffilmark{2},
B. Mobasher\altaffilmark{3},
A. M. Koekemoer\altaffilmark{3},
P. Capak\altaffilmark{4},
Y. Taniguchi\altaffilmark{5},
S. S. Sasaki\altaffilmark{4,5},
H. J. McCracken\altaffilmark{6},
Y. Mellier\altaffilmark{6},
E. Bertin\altaffilmark{6},
H. Aussel\altaffilmark{7,6},
D. B. Sanders\altaffilmark{7},
O. Le Fevre\altaffilmark{8},
O. Ilbert\altaffilmark{7},
M. Salvato\altaffilmark{4},
D. J. Thompson\altaffilmark{16},
J. S. Kartaltepe\altaffilmark{7},
N. Scoville\altaffilmark{4,9},
T. A. Barlow\altaffilmark{4},
K. Forster\altaffilmark{4},
P. G. Friedman\altaffilmark{4},
D. C. Martin\altaffilmark{4},
P. Morrissey\altaffilmark{4},
S. G. Neff\altaffilmark{15},
M. Seibert\altaffilmark{4},
T. Small\altaffilmark{4},
T. K. Wyder\altaffilmark{4},
L. Bianchi\altaffilmark{10},
J. Donas\altaffilmark{8},
T. M. Heckman\altaffilmark{12},
Y.-W. Lee\altaffilmark{11},
B. F. Madore\altaffilmark{14},
B. Milliard\altaffilmark{8},
A. S. Szalay\altaffilmark{12},
B. Y. Welsh\altaffilmark{13},
S. K. Yi \altaffilmark{11}
}

\altaffiltext{*}{Based on observations with the NASA/ESA {\em
Hubble Space Telescope}, obtained at the Space Telescope Science
Institute, which is operated by AURA Inc, under NASA contract NAS
5-26555; and with the NASA {\em Galaxy Evolution Explorer} (GALEX); also based on data collected at : the Subaru Telescope, which is operated by
the National Astronomical Observatory of Japan; Kitt Peak National Observatory, Cerro Tololo Inter-American
Observatory, and the National Optical Astronomy Observatory, which are
operated by the Association of Universities for Research in Astronomy, Inc.
(AURA) under cooperative agreement with the National Science Foundation; 
the Canada-France-Hawaii Telescope with MegaPrime/MegaCam operated as a
joint project by the CFHT Corporation, CEA/DAPNIA, the National Research
Council of Canada, the Canadian Astronomy Data Centre, the Centre National
de la Recherche Scientifique de France, TERAPIX and the University of
Hawaii.}  

\altaffiltext{1}{Department of Astronomy, Columbia University, MC2457,
550 W. 120 St. New York, NY 10027; \email{michel@astro.columbia.edu} \email{ds@astro.columbia.edu}}

\altaffiltext{2}{Department of Physics and Astronomy, University of
California, Los Angeles, CA 90095}

\altaffiltext{3}{Space Telescope Science Institute, 3700 San Martin
Drive, Baltimore, MD 21218}

\altaffiltext{4}{California Institute of Technology, MC 105-24, 1200 East
California Boulevard, Pasadena, CA 91125}

\altaffiltext{5}{Astronomical Institute, Graduate School of Science,
         Tohoku University, Aramaki, Aoba, Sendai 980-8578, Japan}

\altaffiltext{6}{Institut d'Astrophysique de Paris, UMR 7095, 98 bis Boulevard Arago, 75014 Paris, France}

\altaffiltext{7}{Institute for Astronomy, 2680 Woodlawn Dr., University of Hawaii, Honolulu, Hawaii, 96822}

\altaffiltext{8}{Laboratoire d'Astrophysique de Marseille, BP 8, Traverse
du Siphon, 13376 Marseille Cedex 12, France}

\altaffiltext{9}{Visiting Astronomer, Univ. Hawaii, 2680 Woodlawn Dr., Honolulu, HI, 96822}

\altaffiltext{10}{Center for Astrophysical Sciences, The Johns Hopkins
University, 3400 N. Charles St., Baltimore, MD 21218}

\altaffiltext{11}{Center for Space Astrophysics, Yonsei University, Seoul
120-749, Korea}

\altaffiltext{12}{Department of Physics and Astronomy, The Johns Hopkins
University, Homewood Campus, Baltimore, MD 21218}

\altaffiltext{13}{Space Sciences Laboratory, University of California at
Berkeley, 601 Campbell Hall, Berkeley, CA 94720}

\altaffiltext{14}{Observatories of the Carnegie Institution of Washington,
813 Santa Barbara St., Pasadena, CA 91101}

\altaffiltext{15}{Laboratory for Astronomy and Solar Physics, NASA Goddard
Space Flight Center, Greenbelt, MD 20771}

\altaffiltext{16}{Caltech Optical Observatories, MS 320-47, California Institute of Technology, Pasadena, CA 91125}

\begin{abstract}
We present a study of the morphological nature of redshift $z \sim 0.7$ star-forming galaxies using a combination of HST/ACS, GALEX and ground-based images of the COSMOS field.   Our sample consists of 8,146 galaxies, 5,777 of which are detected in the GALEX near-ultraviolet band $(2310\AA \mbox{ or} \sim 1360\AA \mbox{ rest-frame})$ down to a limiting magnitude of 25.5 (AB), and all of which have a brightness of $F814W \mbox{ (HST) } < 23 \mbox{ mag}$ and photometric redshifts in the range $0.55 <  z < 0.8$.  We make use of the UV to estimate star formation rates, correcting for the effect of dust using the UV-slope, and of the ground-based mutli-band data to calculate masses.  For all galaxies in our sample, we compute, from the ACS F814W images, their concentration (C), asymmetry (A) and clumpiness (S)  as well as their Gini coefficient (G) and the second moment of the brightest 20\% of their light (M20).  We observe a bimodality in the galaxy population in asymmetry and in clumpiness, though the separation is most evident when either of those parameters is combined with a concentration-like parameter (C, G or M20).  We further show that this morphological bimodality has a strong correspondence with the FUV - $g$ color bimodality, implying that UV-optical color predominantly evolves concurrently with morphology.  We observe many of the most star-forming galaxies to have morphologies approaching that of early-type galaxies, and interpret this as evidence that strong starburst events are linked to bulge growth and constitute a process through which galaxies can be brought from the blue to the red sequence while simultaneously modifying their morphology accordingly.  We conclude that the red sequence has continued growing at $z \lesssim 0.7$.  We also observe $z \sim 0.7$ galaxies to have physical properties similar to that of local galaxies, except for higher star formation rates.  Whence we infer that the dimming of star-forming galaxies is responsible for most of the evolution in star formation rate density since that redshift,  although our data are also consistent with a mild number evolution.
\end{abstract}


\section{INTRODUCTION}

\setcounter{footnote}{16}

Star formation has been in decline, in the Universe, for the past 8 billion years.  This discovery, manifested in the now well-known Madau diagram \citep{Lilly96,Madau96},  has been a remarkable culmination of last decade's research in galaxy evolution.  The details of this decline remain, however, surprisingly elusive:  the reason lying in the complex nature of star formation itself.   Fuel exhaustion ({\em in situ} gas consumption), reduction in merger rate and environmental effects could all share the responsibility.  As many physical characteristics of galaxies correlate with their Hubble type \citep[e.g.][]{Kennicutt98}, their classification within redshift surveys has been a natural segue in the investigation of the decline of star formation.  

Color, spectral class or morphology are among the most commonly used criteria to separate galaxies and were quickly applied by investigators.  For example, \citet{Lilly95} found that the luminosity density of blue galaxies brightens substantially up to redshift $z \sim 1$, whereas that of red galaxies does not, a result confirmed by today's much larger samples \citep{Faber05}.  \citet{Ellis96} divided galaxies in the Autofib/LDSS sample according to their [OII] equivalent width and found a strong evolution in the volume density of moderate to low-luminosity objects with strong [OII] emission while \citet{Heyl97} divided the same sample into spectral types and found that late-type spirals were the ones dominating the evolution of the blue luminosity function.

Attempts at resolving distant star-forming galaxies into morphological types date back to the problem of faint blue galaxies \citep[see][for a review]{Ellis97} for which \citet{Brinchmann98} concluded, using HST pointings at locations in the CFRS \citep{Lilly95} and LDSS \citep{Ellis96} fields, that peculiar galaxies, identified to be responsible for the faint blue excess \citep{Griffiths94}, were also the main cause for the rapid evolution of the blue luminosity function observed in those redshift surveys.  Recently, however, \citet{Wolf05}, with a sample of 1483 galaxies at $z \sim 0.7$ extracted from GEMS \citep{Rix04}, found spiral galaxies to actually dominate the overall UV($2800\AA$) luminosity at that redshift (though with irregular galaxies still being prevalent at faint magnitudes), implying that their fading, accompanied with a similar migration to lower UV-luminosities in irregulars, must lead the decline of star formation density, rather than a decrease in merging rate.  {\em Spitzer} $24 \mu m$ observations of that same sample show that most of the IR-emission associated with dust-reprocessed UV-light from young stars, emission that declines with redshift even faster than that of the escaping UV \citep{LeFloc'h05}, is also dominated by spiral galaxies \citep{Bell05}.  Moreover, \citet{Melbourne05} compared LIRG morphologies at redshifts $0.1<z<1$ and confirmed that high-redshift ($z>0.5$) LIRGs are dominated by spirals unlike low-redshift ones ($z<0.5$) which are mostly peculiars.  They interpret that as a depletion of gas supply causing spirals to fall to sub-LIRG levels of star formation, while peculiar morphologies, characteristic of mergers, continue to experience strong bursts of star formation.  Along the same lines, \citet{Menanteau06}, using the parallel NICMOS observations of the UDF, also showed spirals to dominate $\rho_{SFR}$ at all redshifts up to $z \gtrsim 1$, at which point irregular/peculiar galaxies, which show the sharpest rise, become equally important.  These many evidences suggest that spirals play a crucial role in the last 8 Gyr evolution of galaxies.

Meanwhile, a wealth of information about low-redshift galaxies has also emerged.  In particular, large surveys such as SDSS \citep{York00} and 2dF Galaxy Redshift Survey \citep[e.g.][]{Madgwick02} have really propelled statistical studies of galaxy properties.  Although trends of color vs. morphology have been known for a long time \citep{deVaucouleurs61}, \citet{Strateva01} showed that the color distribution of galaxies was not smooth, but doubly-peaked with early-types dominating the red population and late-types the blue one.  \citet{Kauffmann03a} also observed this bimodality in the $D_{n}(4000)$ index, indicative of a division in the population between galaxies dominated by an old stellar population and ones that experienced recent episodes of star formation.  \citet{Kauffmann03b} further showed that this separation occurs at $M_{*} \sim 3 \times 10^{10} M_{\odot}$ and that lower-mass galaxies besides having young stellar populations also have disk-like structural parameters, whereas higher-mass ones have old stellar populations and bulge-like structural parameters.  \citet{Brinchmann04} reinforced these conclusions by directly calculating specific star formation rates (SFR per unit mass) for $\sim 150,000$ galaxies, and observed the same divide of high-sSFR and low-sSFR with age, mass and structural parameters/Hubble type.  \citet{Blanton03b} labeled these two populations the red and the blue sequences.  More recently, \citet{Bell04} found the same color bimodality at all redshifts up to $z \sim 1$ using the COMBO-17 sample \citep{Wolf03}.

With COSMOS \citep{Scoville07}, it is now possible to expand statistical analyses of higher redshift galaxies to levels that allow comparison with SDSS or 2dF.  In this paper, we combine HST/ACS and GALEX coverage of the COSMOS field \citep{Koekemoer07,Schiminovich07} to study the morphological properties of star-forming galaxies at $z \sim 0.7$.  In view of the streneous task of classifying large numbers of objects by eye and of the subjectivity it carries, for this study as well as for future comparison we chose to follow the path of automated classification.  \citet{Abraham96,Abraham96b} demonstrated the usefulness of such an approach by using concentration and asymmetry measurements to classify galaxies in the MDS and HDF according to their location in the C-A plane.  In parallel, \citet{Odewahn96} used a neural-network code on the HDF sample and arrived at results similar to \citet{Abraham96b}.  Following along this path, people have devised and applied several ingenious algorithms for morphological classification, the most popular being the Sersic index \citep{Sersic68}, bulge-to-disk decomposition \citep{Simard98, Peng02}, and shapelet decomposition \citep{Refregier03}.  These, however, all require fitting objects to a set of parameters or functions.  In this paper, we preferred to follow the purely mensurational approach of \citet{Abraham96}, and expanding it by adding measurements of clumpiness \citep{Conselice03} as well as of the recently developed Gini and M20 coefficients \citep{Lotz04,Abraham03} to use as our morphological parameters, and classification criteria.

In this paper, we present morphological characteristics of the galaxy population at $z \sim 0.7$, and study their relation to physical parameters.  We also compare both their morphological and physical properties with those of low-redshift galaxy samples in the literature.  We focus most of our attention on the relation between star formation rate and morphology, and interpret our results in the framework of galaxy evolution.  We discuss implications for evolution scenarios since $z \sim 0.7$ in the context of the literature, with an emphasis on blue to red sequence evolution.  The paper is organized as such: we first shortly describe, in section 2, the COSMOS survey, its data, as well as the GALEX observations and data, before discussing our sample selection.  We outline our approach to morphological analysis in section 3, present the results of our investigation in section 4, and discuss their interpretation and implications in section 5.

\section{OBSERVATIONS and DATA}

\subsection{HST/ACS Observations}

We make use of the full HST coverage of the COSMOS field, which consists of 542 HST/ACS images with depth of $I < 27 \mbox{ mag (AB,}10\sigma \mbox{ point source})$, $0.09"$ FWHM resolution (with $0.05"$ pixels) and whose mosaic spans an area of $2 \deg^{2}$.  An overview of the COSMOS project is given in \citet{Scoville07} with detail description of the ACS  observations and data reduction in \citet{Scoville07b} and \citet{Koekemoer07} respectively.

\subsection{Ground-based Observations and Catalog}

Ground-based follow-up observations have been performed using the CFHT ({\em u*} and {\em i} bands), Subaru/SuprimeCam (BV{\em griz}), Kitt Peak/Flamingos (K-band) and CTIO (also K-band) telescopes, providing deep coverage, with typical limiting magnitudes of 27 (AB, $3\sigma$), of the field from the $u$ to $z$ bands ($m_{limit}^{z} = 25.8$), as well as shallower imaging in the K-band ($m_{limit}^{K} = 21.6$).  Details of the ground-based observations and data reduction are presented in \citet{Capak07} and \citet{Taniguchi07}.  A multi-wavelength photometric catalog \citep{Capak07} was generated using SExtractor \citep{Bertin96}, with the $i$-band as the selection wavelength.  We further performed SED fitting of this multi-band data and calculated photometric redshifts for galaxies with $i < 25$ mag (AB) \citep{Mobasher07}.  Our photometric redshifts have an rms of $\sigma((z_{phot}-z_{spec})/(1+z_{spec})) = 0.031$ with 2\% outliers.  We make use of these photometric redshifts in selecting our sample.

\subsection{GALEX  Observations}

We used GALEX \citep{Martin05}, which has a circular field-of-view of $1.2^{\circ}$ in diameter, to observe the COSMOS region in ultraviolet light with four pointings of $\sim 50$ ks each.  These observations, performed as part of the GALEX Deep Imaging Survey, reach a limiting magnitude of $\sim 25.5$ mag (AB) in the near-ultraviolet band (NUV).  GALEX has a resolution of 5.6" in the NUV which corresponds to $ \sim 40$ kpc at $ z \sim 0.7$, larger than the typical size of galaxies.  The full details of the observations can be found in \citet{Schiminovich07}.

Since standard pipeline processing of deep fields can sometimes blend two objects into a single detection, we employed a different method of source extraction.  As the vast majority of our sources appear unresolved to GALEX, we decided to use the DAOPHOT software \citep{Stetson92} to measure photometry.  We further made use of the ground-based COSMOS catalog to feed DAOPHOT with position priors.  Because of small astrometric offsets, our first step was to align the priors with each GALEX image.  We ran the {\em phot} routine (a routine that performs aperture photometry) with {\em centroid} recentering of the objects.  We then fit the center of the distribution of shifts in the x and y-directions as well as in the angle of rotation $\theta$ around the center of the pointing, and applied the mean shift to all positions obtained from the astrometry.  This small ($< 1$ pix) constant uniform shift does a good job at realigning position priors with objects, and is therefore the only astrometric correction we applied.

We then followed the standard DAOPHOT procedures of running {\em phot} (performing aperture photometry), {\em psf} (modeling the psf) and {\em allstar} (performing psf-fitting photometry) to obtain UV-fluxes for our objects.  This time we did not allow for recentering in the {\em phot} procedure, but we did in {\em allstar}, since, by looking at the residuals, we found psf-fitting to be much better when recentering was allowed.  The drawback, however, is that priors located in regions with no apparent UV were often moved, in the process of recentering, to fit a neighboring object.  The measurements were therefore rematched to objects in the original catalog that were located closest, but no farther than $3"$, to the measured GALEX positions.  Lastly, we created masks to eliminate various artifacts, as well as a few very bright stars, in our four NUV images, and nulled all detections found inside masked regions.

\subsection{Sample Selection}

\begin{figure}[hbtp]
\begin{center}
\includegraphics[width=3.4in]{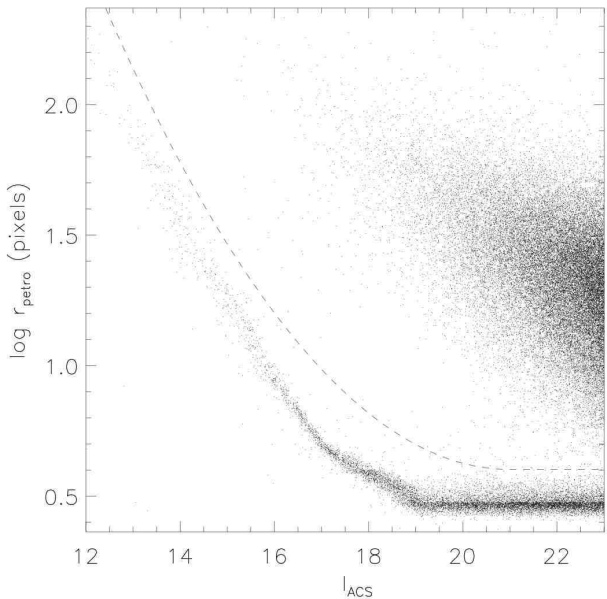}
\caption{Petrosian radius vs $I_{ACS}$-band magnitude for all objects in our morphological sample.  The dashed-line represents our star/galaxy separation, stars lying below the line and galaxies above.}
\label{fig:star-gal_sep}
\end{center}
\end{figure}

We aimed at extracting a sample in a narrow range of redshifts around $z\sim 0.7$ bright enough to study morphology in the ACS images.  The choice of redshift $0.7$ is convenient in that the observed NUV-band roughly corresponds to the ($z = 0.1$)-frame FUV.  This minimizes K-corrections and allows for easy comparison with local samples such as SDSS.  A narrow redshift range further allows us to obviate the need for morphological K-corrections.  The selection procedure we employed is the following:  we first ran our morphological analysis exclusively on objects with $I \mbox{ mag } < 23$ as morphological parameters become less reliable for fainter objects.  We then removed all objects with (petrosian) radii smaller than 0.2", since they are below our resolution limit.  As those objects are mostly stars and QSO's (figure~\ref{fig:star-gal_sep}), 
this cut does not introduce any bias in our sample.  Our magnitude cut does, however, progressively bias us towards higher surface brightness objects as we move to smaller radii.  This effect shows up later in some of our analysis, and is discussed in context.  We further cleaned our sample of stars (and stellar-like object) by applying a cut in the $r_{petro} - I_{ACS}$ plane (these parameters are described in the next section), where stars and galaxies segregate unmistakably (figure~\ref{fig:star-gal_sep}), before proceeding to remove objects that, after visual inspection, were found to be false detections or that showed various problems with their segmentation (we describe our segmentation technique in section 3.1).  These fews steps not only clean our sample of stars, but also, at the same time, of most but the weakest AGNs.  We lastly selected for this study only objects with photometric redshifts in the range $0.55 <  z \leq 0.8$.  This redshift bin width translates into a difference of luminosity of 0.8 magnitudes for objects with the same brightness located on both ends of the redshift range.  In the end, our sample contains $8,146$ galaxies, $5,777$ of which are detected in the NUV with GALEX.  We thus detect, in the UV, about $70\%$ of objects with $I\mbox{ mag } < 23$ and redshift $z\sim 0.7$.  Throughout this paper, we sometimes utilize, where appropriate, our {\em UV-detected} sample only, but otherwise normally refer to our {\em full} sample of $8,146$ galaxies.

\subsection{Star Formation Rates and Masses}

We performed our own SED analysis on the combined GALEX + ground-based photometric data using the KCORRECT software \citep{Blanton03a}, and extracted from it K-correction estimates for each galaxy.  We chose such an approach because KCORRECT is designed to extract the most physically realizable SED by using linear combinations of four spectra that are characteristic of physical states of galaxies, from intensely starbursting to quiescent.  We applied the K-corrections to $u*$ and NUV-band photometry to obtain restframe FUV and NUV absolute magnitudes from which we derived a UV-slope which, given the relation between $\beta$ (the UV-slope) and $A_{FUV}$ (the FUV attenuation) \citep{Seibert05}, provides us with a dust correction factor.\footnote{Some observations \citep{Seibert05,Cortese06} suggest that the $A_{FUV} - \beta$ relation is steeper in starburst galaxies than in normal galaxies.  Since we are using the relation for normal galaxies, it is possible that, for starbursts, our star formation rates are slightly underestimated.  This, however, would not affect the qualitative behavior of star formation rate in relation to other properties, which is what we focus on in this paper.}   We then converted the corrected FUV-luminosities to star formation rates using the \citet{Kennicutt98} relation between star formation rate and UV-continuum luminosity:
\begin{equation}
\mbox{SFR}(M_{\odot} \mbox{ year}^{-1}) = 1.4 \times 10^{-28} L_{\nu, UV}(\mbox{ergs s}^{-1} \mbox{ Hz}^{-1})
\label{eq:uvtosfr}
\end{equation}
For objects without NUV counterparts, restframe FUV magnitudes have been derived directly from the fitted SEDs.  Furthermore, because of the degeneracy between age and dust in red galaxies, and because of the uncertainty in FUV magnitudes derived for objects with no GALEX detection, we decided not to use $\beta$ as a proxy for dust attenuation for these galaxies, but instead, to simply apply a moderate constant dust correction of $+0.5$ in log SFR (which is equivalent to a $A_{FUV}$ of 1.25 or an E(B-V) of 0.151).  This is a reasonable correction for early-type galaxies which constitute most of these objects.  On the other hand, this method completely misses the most heavily obscured galaxies, such as could be some ULIRGs, though these are far less common.  The upcoming {\em Spitzer} data release for the COSMOS field will be extremely helpful in the study of these objects.  For now, we need to leave those with UV-fluxes below our detection limit behind,  i.e. with star formation rate estimates in the range of quiescent galaxies, much below their true value.  Because of the discrepancy in the quality of our measurements between UV-detected and non-UV-detected objects discussed above, we clearly differentiate the two populations in our plots and analysis.

After conversion to a star formation rate, our limiting magnitude of $m_{NUV} = 25.5$ corresponds to log SFR = 0.11 or a star formation rate of about $1 \mbox{ M}_{\odot} \mbox{ yr}^{-1}$ for a $z = 0.7$ galaxy with an $A_{FUV}$ of 1.25.  Throughout this paper, we thus also refer to objects in our UV-detected sample as star-forming galaxies.

We also applied our derived K-corrections to obtain restframe $B$ and $V$-band absolute magnitudes and used the \citet{Bell01} relation between $B-V$ color and the ratio of mass to $V$-band luminosity 
\begin{equation}
\log (M_{*}/L_{V}) = -0.734 + 1.404 \times (B-V)
\end{equation}
with a scaled Salpeter IMF to calculate masses for all objects in our sample.  The scale Salpeter IMF \citep{Bell01} has a shallower slope at low masses, similar to the Chabrier IMF \citep{Chabrier03}.
This method is accurate to about 0.1-0.2 dex with most uncertainties coming from bursts of star formation, dust and uncertainties in the \citet{Bell01} models.

\section{MORPHOLOGICAL ANALYSIS}

Given the large nature of our sample, it is important to use both an automated and consistent morphological classification scheme.  We therefore chose to follow the work of \citet{Abraham96}, \citet{Conselice00}, \citet{Conselice03}, \citet{Abraham03} and \citet{Lotz04} and use their non-parametric approaches, thus computing Concentration(C), Asymmetry(A) and Clumpiness(S) parameters \citep{Conselice03}, as well as the Gini coefficient(G) and second order moment of the distribution of the brightest 20\% of the light(M20) \citep{Lotz04} for all objects in our sample.  We briefly describe these parameters in the following sections, but refer the reader to the papers cited above for a full description.

\subsection{Size and Segmentation}

Before measuring any morphological parameter, one needs to assign a region of the image to every galaxy.  The standard approach \citep{Conselice03} has been to take a circular aperture of radius $1.5 \times r_{petro}$, where $r_{petro}$, the Petrosian radius, is the radius at which $\eta (r)  = 0.2$, and where $\eta (r)$ is defined as
\begin{equation}
\eta (r) = \frac{\mu (r)}{\mu (<r)}
\end{equation}
that is the ratio of the surface brightness at a given radius to the mean surface brightness within that radius.  Over a fixed surface brightness cut, this method has the advantage of being far less affected by surface brightness dimming.

The Gini coefficient and the second order moment of the light, however, require a full segmentation of every object. \citet{Lotz04} use the isophote of surface brightness $\mu = \mu (r_{petro})$ on a smoothed version of the image, the smoothing kernel being a gaussian of $\sigma = 0.2 \times r_{petro}$, as the boundary of their galaxies.  We followed the same prescription, though with a tophat smoothing kernel of diameter $0.3 \times r_{petro}$.  We also imposed a minimum surface brightness of $\mu_{min} = 0.6 \times \sigma_{background}$ (after background subtraction).  By summing all the flux within our segmentation maps, we were able to estimate the total $I_{ACS}$-band flux for each galaxy, which we then used to normalize all of our morphological and size parameters.  Unlike \citet{Conselice03} who used thumbnail images, our algorithm extracts objects from larger images, and we therefore decided to also extend the use of segmentation maps to the computation of asymmetry and clumpiness since, compared to circular apertures, they are less likely to pick up light from neighboring objects.

\begin{figure}[htbp]
\begin{center}
\includegraphics[width=3.4in]{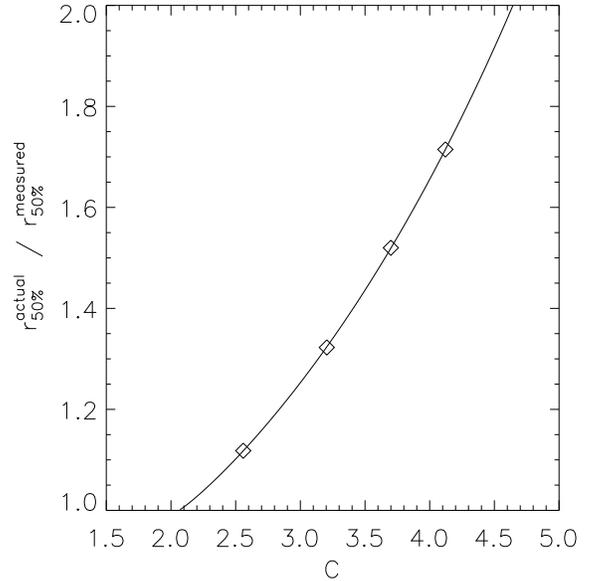}
\caption{Actual 50\%-radius over the measured 50\%-radius as of function of measured concentration, for theoretical S\'ersic profiles of integer indices 1 to 4.  The line is a fit through the four calculated points and represents the size correction factor we applied to our measured values of $r_{50\%}$ in order to recover actual values of $r_{50\%}$ for our objects.  We apply a minimum and maximum correction factor of 1 and 2, respectively.}
\label{fig:size_corr}
\end{center}
\end{figure}

Apart from $r_{petro}$, we also make use of three other size measurements:  $r_{20\%}$, $r_{50\%}$ and $r_{80\%}$.  These represent the radii encompassing respectively 20, 50 and 80\% of the total $I_{ACS}$-flux of the galaxy, and are obtained by summing the flux inside a circular aperture of expanding radius until the respective percentages of the light are attained.  We utilize $r_{20\%}$ and $r_{80\%}$ in the computation of concentration, and we use $r_{50\%}$ (converted to a physical scale) as our parameter for size.  Because our value of $r_{50\%}$ is measured with respect to the total flux inside the Petrosian radius, we systematically underestimate its actual value.  Therefore, in order to attempt to recover to true size of our objects, we multiplied our values of $r_{50\%}$ by a correction factor.  This correction factor was calculated by, first comparing true values of $r_{50\%}$ for theoretical S\'{e}rsic profiles \citep{Sersic68} with S\'{e}rsic indices of 1,2,3 and 4, to the radius at 50\% of the flux inside their Petrosian radius, then calculating the concentration values for those four profiles (there exists a one-to-one monotonic relation between S\'{e}rsic index and concentration), and finally obtaining a fit for $r_{50\%}^{actual} / r_{50\%}^{measured}$ as a function of concentration.  The result is shown in figure~\ref{fig:size_corr}.  The points represent the theoretical calculations for S\'{e}rsic profiles of S\'{e}rsic indices 1,2,3 and 4, and the line represents the fit to the relation.  We also limit the correction factor to 2.0, hence all objects with $C > 4.65$ simply have their $r_{50\%}$ doubled.

\subsection{Concentration}

Concentration is defined by the ratio of the radius containing 80\% ($r_{80\%}$) to the radius containing 20\% ($r_{20\%}$) of the total light.
\begin{equation}
C=5\times \log \frac{r_{80\%}}{r_{20\%}}
\end{equation}

\subsection{Asymmetry}

Asymmetry is calculated by comparing an object with an image of itself rotated by $180^{\circ}$.  It is therefore crucial to know the object's center, which becomes the pivot point.  Our approach in determining centers mostly follows the one described in \citet{Conselice00}.  We first pick the brightest pixel, after some smoothing, as a first estimate of the center.  We then refine it by calculating asymmetries within a 4-pixel radius aperture successively centered on each of the 9 points of a $3 \times 3$ grid at that initial center. We proceed to the lowest asymmetry point, refining our mesh to sub-pixel level by interpolation until the difference in asymmetries between the lowest and second to lowest points is less  than or equal to 0.001 or 20 iterations have been reached.  Contrary to \citet{Conselice00} we only use a 4-pixel radius aperture, as opposed to full-aperture, to minimize our asymmetry, since our goal is really to find the center of symmetry of the bulge.  Minimizing global asymmetry would generally give us a center closer to the center of light of the system, which, in the case of highly peculiar galaxies or mergers, could be very far from the bulge center.   Although that approach is just as valid, we found the first one to be a better discriminant of interacting systems.

We implemented two other minor changes in the asymmetry algorithm.  Both are mostly procedural, help reduce the scatter, but also tend to produce values of A that are higher than that of the standard algorithm \citep{Conselice00}, though, we reckon, more accurate.  One modification is the use of segmentation maps instead of circular apertures as mentioned above.  More precisely, the procedure involves symmetrizing the maps first, and then applying them to the difference image obtained by subtracting the rotated image from its original.  This causes interacting systems to be fully included into the segmentation rather than only the part of which falls within a certain circular apperture.  The second modification we implemented concerns the way we estimate the effect of the background.  \citep{Conselice00} use a nearby empty region of space to calculate the asymmetry of the background and then subtract that asymmetry from the original value.  We, on the other hand, estimate the effect of the background by calculating a second asymmetry value ($A'$) from a convolved version of the object, with the following 5-point average convolution
\begin{equation}
f'_{i,j} = \frac{1}{5}(f_{i,j}+f_{i+1,j}+f_{i-1,j}+f_{i,j+1}+f_{i,j-1})
\end{equation}
where $f_{i,j}$ represents the flux at the $(i,j)$ pixel of the image.  If we assume that the intrinsic asymmetry of the light does not change in the weakly convolved version, and there is evidence for such an assumption to hold with even bigger convolution kernels \citep{Conselice03}, the difference between the two asymmetry values must be entirely due to background, and since the standard deviation of the background in the smoothed image is reduced by a factor $\sqrt{5}$ from its unsmoothed version, we have
\begin{eqnarray}
A_{intrinsic}  & = & A - A_{background} \\
& = & A' - A'_{background}\\
& = & A' - \frac{A_{background}}{\sqrt{5}}
\end{eqnarray}
This implies that we can estimate the amount of asymmetry due to random fluctuations in the background from the following formula:
\begin{equation}
A_{background} = \frac{A - A'}{1-1/\sqrt{5}}
\end{equation}
and subtract it from our asymmetry measurement to obtain the intrinsic asymmetry of the object.  This prevents us from subtracting background asymmetry from regions where it is due to intrinsic differences between opposite parts of the galaxy.

\subsection{Clumpiness}

Clumpiness is calculated by subtracting from an image a blurred version of itself.   The blurred version is obtained by convolving the image with a circular tophat filter of diameter equal to $0.3 \times r_{petro}$.  After subtraction, only positive values are retained and summed.  Straight forward application of this procedure almost always retains significant flux in the center, where bulges sharply peak, and must be corrected for.  \citet{Conselice03} chose to simply blank the region inside one filter radius.  We opted for a similar but slightly different approach.  We decreased the size of our convolution kernel for points inside two filter radii.  In that regime, we set the filter radius to one half of the distance to the center.  It reaches zero in the center, with the 9 central pixels not being smoothed at all, therefore always subtracting out.  This allowed us to minimize the contribution of the central peak to the clumpiness value while still picking out bright clumps or bars near the center.  In most cases though, this slight modification did not yield different results from the standard approach.  As for the effect of background on clumpiness, we corrected for it using the standard method \citep{Conselice03}.  We also make use, here, of segmentation maps for summing over galaxy pixels in the difference image.

\subsection{Gini}

The Gini coefficient \citep{Abraham03} measures the inequality of the distribution of flux among the pixels associated to a galaxy.  Its possible range of values goes from 0, in the case where all the pixels would have the same intensity, i.e. complete equality among pixels, to 1, in the case where all the flux of a galaxy would be contained in a single pixel.  In general, it can be defined as the ratio of the area between the Lorenz curve and the curve of uniform equality to the area under the curve of uniform equality.  The Lorenz curve, $L(p)$, is in turn defined as the curve representing the proportion of the total flux contained in the dimmest $p$ fraction of pixels.  The Gini coefficient is thus somewhat analogous to the concentration parameter, except that it is not measured with respect to a specified center.  It requires, on the other hand, objects to be segmented.  In other words, a boundary needs to be drawn inside which pixels are assigned to the object.  As described above (¤3.1), we follow the prescription of \citet{Lotz04} and use the isophote of surface brightness $\mu (r_{petro})$ in the convolved image as our boundary.  We then calculated Gini coefficients for our objects by summing the values of the segmented pixels in the following way:
\begin{equation}
G = \frac{1}{\left| \bar{X} \right| n (n-1)} \sum_{i}^{n} (2i - n -1) \left| X_{i} \right|
\end{equation}
where the $n$ pixels are first sorted from dimmest to brightest (in absolute value), and $X_{i}$ represents the flux of the $i^{th}$ pixel.  This method is equivalent to the definition given above, the absolute values making it further more robust to background noise \citep{Lotz04}.

\subsection{M20}

\begin{figure*}[thbp]
\begin{center}
\includegraphics[scale=0.9]{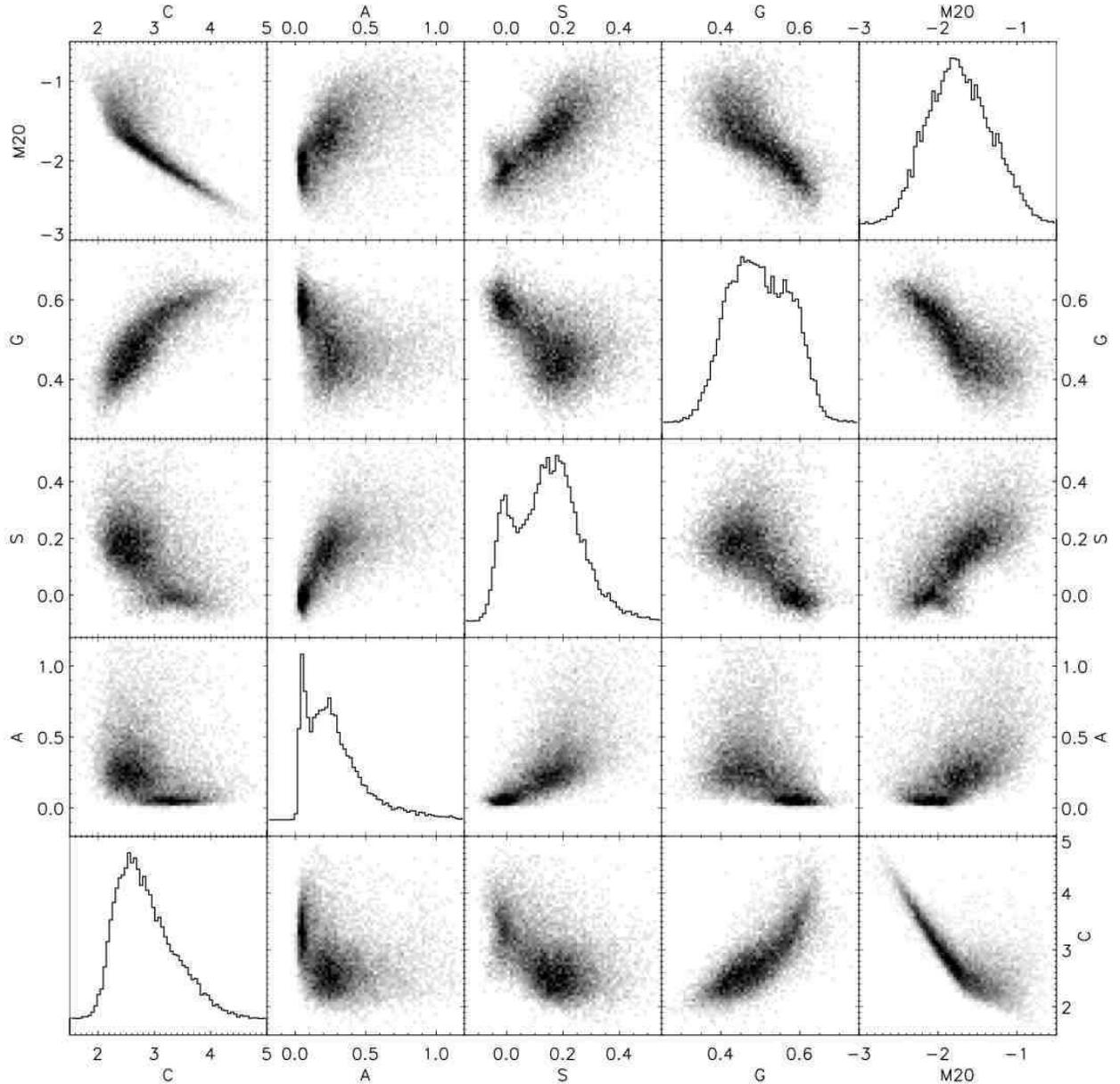}
\caption{The distribution of our full sample in concentration(C), asymmetry(A), clumpiness(S), Gini coefficient(G) and second moment of the brightest 20\% of the light(M20).}
\label{fig:casgm}
\end{center}
\end{figure*}

M20 stands for the normalized second order moment of the brightest 20\% of the galaxy's flux.  It is best described mathematically as:

\begin{equation}
M_{20} \equiv \log10\left ( \frac{\sum_{i} f_{i} \cdot r_{i}^{2}}{M_{tot}}  \right )
\end{equation}

\noindent
where $f_{i}$ and $r_{i}$ represent the flux and distance from the center of the $i^{th}$ pixel respectively, and where the sum is performed by adding pixels in decreasing order of brightness (starting with the brightest one) until $\sum_{i} f_{i}$ reaches 20\% of the total flux.  $M_{tot}$ in this equation is simply the second-order moment summed over all pixels.  M20 is thus like an inverse concentration for galaxies whose profile declines monotonically and isotropically.  In those cases, the brightest 20\% of the flux pixels is equivalent to the region enclosed by the 20\% of the flux radius, and $r_{80\%}$ and $r_{petro}$ then also follow a simple relation.  However, M20 is much more strongly influenced by bright clumps in the outskirts of galaxies than is concentration.  Therefore, whereas concentration can sometimes be thought as a bulge-to-disk ratio, M20 diverges from that concept in cases where extended non-axisymmetric light becomes important, as is the case in mergers and certain disks.

\section{RESULTS}
 
\subsection{Morphology of $z \sim 0.7$ Galaxies}

Figure~\ref{fig:casgm} shows the distribution of our full sample in our five morphological parameters.  The distribution displays a clear morphological bimodality, analogous to the color bimodality observed by \citet{Bell04} at those redshifts.  This bimodality appears in both asymmetry and clumpiness parameters, but becomes most apparent in a 2-D distribution when either of these is combined with a concentration-like parameter (concentration, Gini or M20).  This implies that galaxies tend to be either spheroidal, in which case they have nearly zero asymmetry and clumpiness, or dominated by disks, which exhibit typical values of asymmetry and clumpiness around $A=0.25$ and $S=0.17$.  Intermediate morphologies in which a visible disk still exists albeit being supplanted in importance by the bulge, such as in S0's and Sa's, are encountered less frequently.

\begin{figure*}[thbp]
\begin{center}
\includegraphics{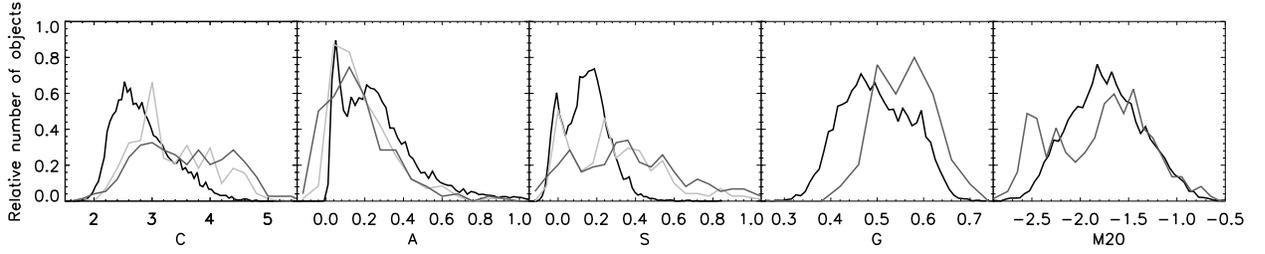}
\caption{Histograms of the relative number of objects in our sample as a function of each of our morphological parameters, compared to the distribution in the samples of \citet{Conselice03} (light grey line) and \citet{Lotz04} (dark grey line).}
\label{fig:casgm-local}
\end{center}
\end{figure*}

Although clumpiness and asymmetry appear to behave in very similar ways and even to actually correlate with one another, they are not completely degenerate.  For example, the galaxies with the highest ratios of $S/A$ are almost exclusively edge-on galaxies, whereas the ones with the lowest ratios often have a bright compact center with a long but faint and smooth tail, or cloud, extending on one side (tadpole galaxies [\citeauthor{Griffiths94}, \citeyear{Griffiths94}] would fall in that category), thus boosting asymmetry, but not clumpiness.  On the other hand, it is true that elliptical galaxies will have both near-zero asymmetry and clumpiness.

\begin{figure*}[bhtp]
\begin{center}
\subfigure[]{\includegraphics[scale=0.75]{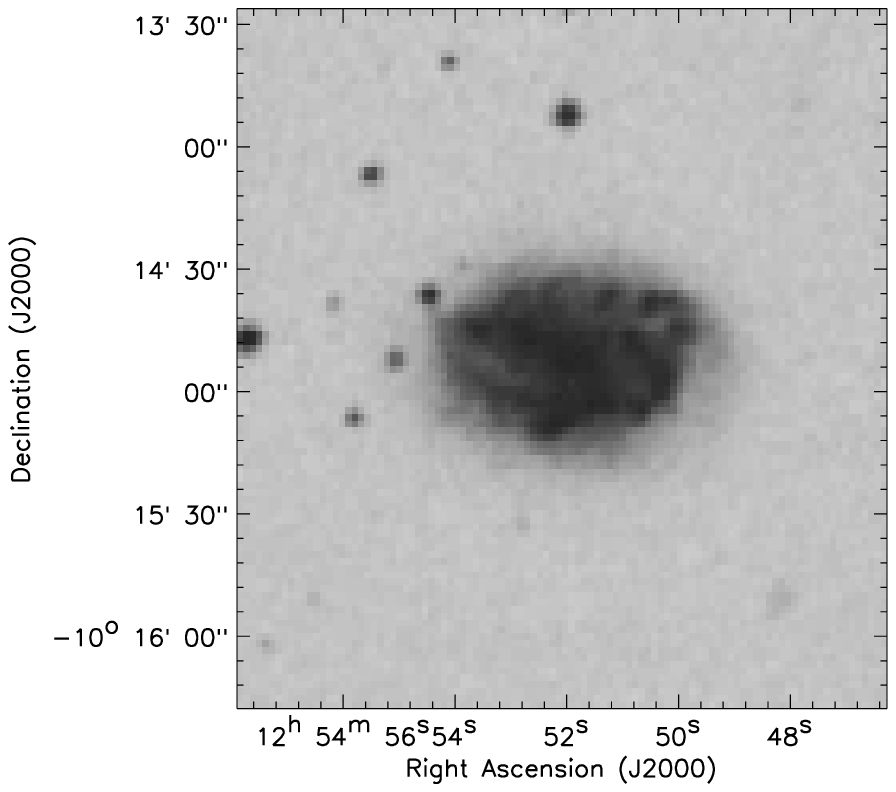}}\quad
\subfigure[]{\includegraphics[scale=0.75]{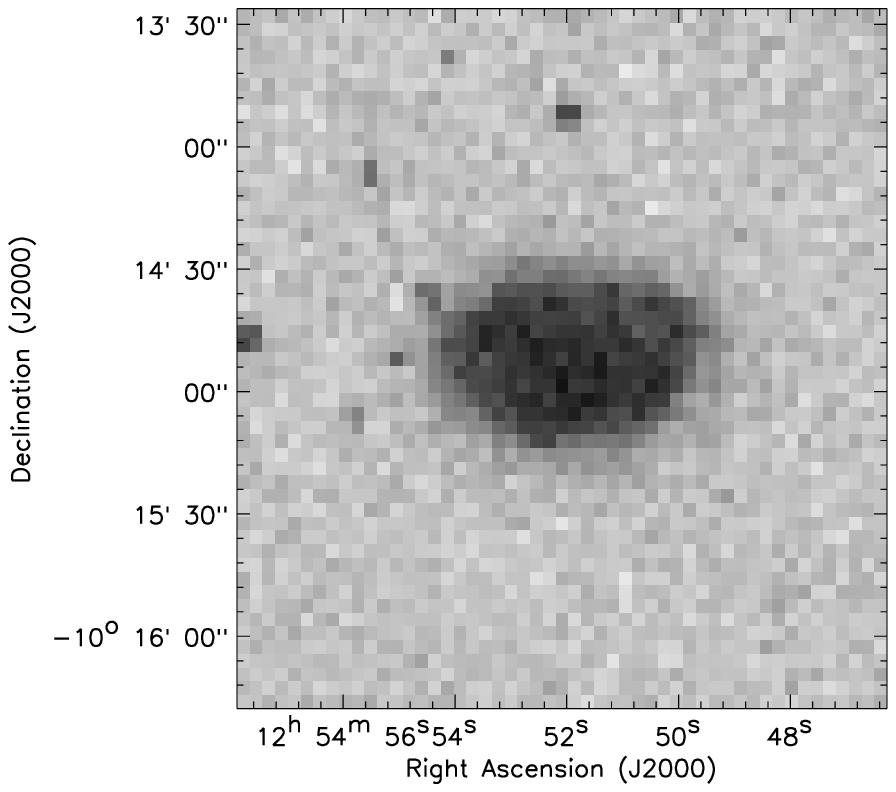}}\\
\subfigure[]{\includegraphics[scale=0.75]{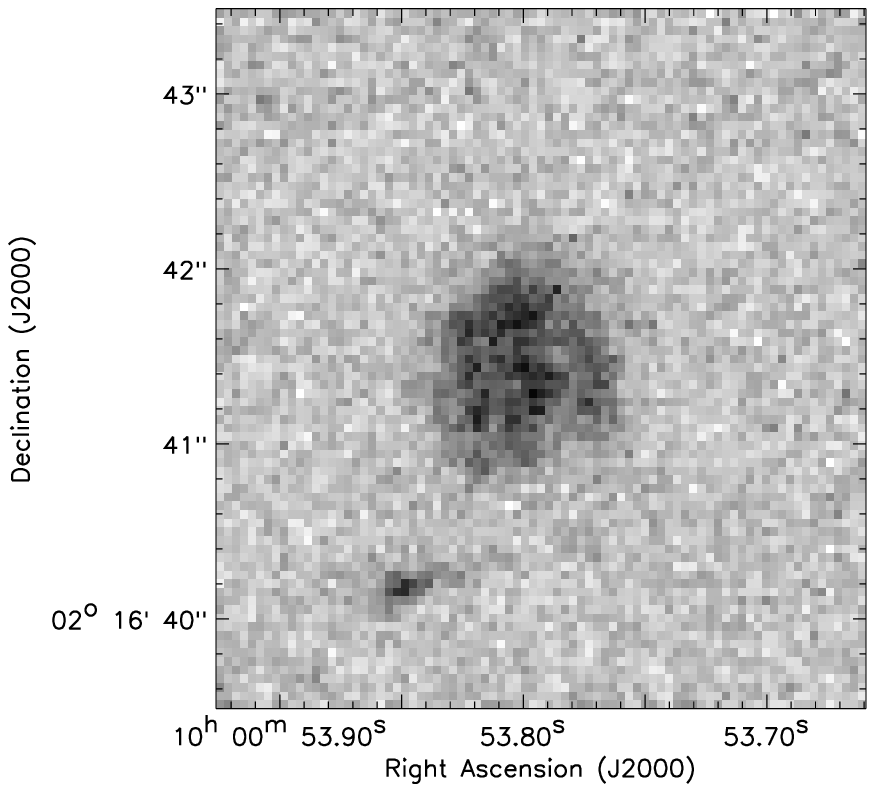}}\quad
\subfigure[]{\includegraphics[scale=0.75]{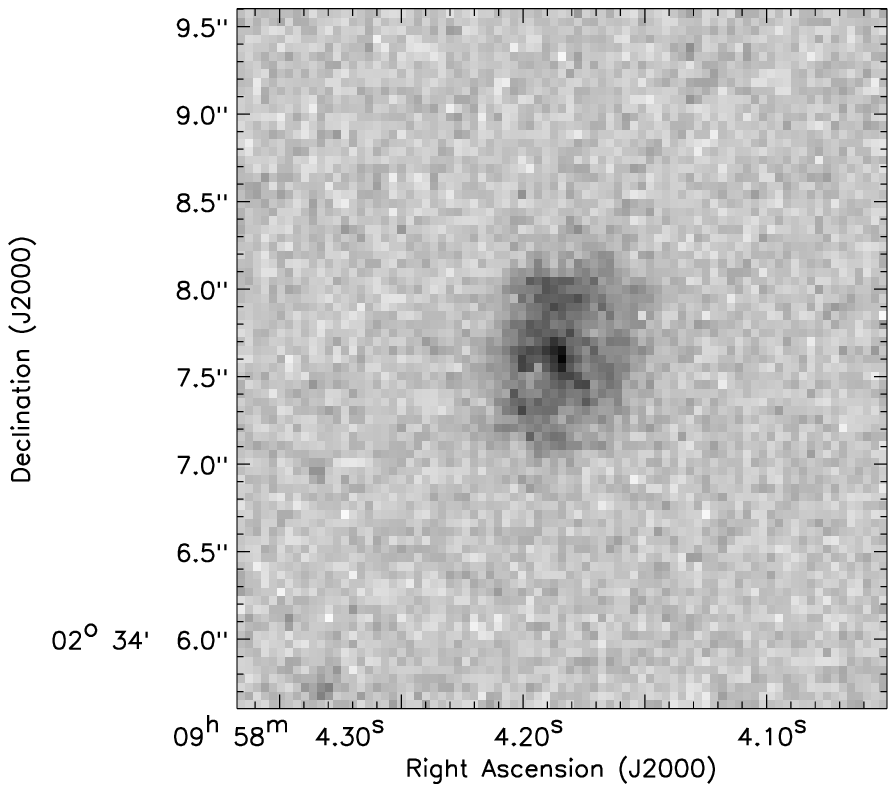}}
\caption{a) Sd galaxy NGC 4790.  NGC 4790 has a Gini coefficient of 0.4 and is 19 Mpc away.  b) Degraded image of NGC 4790.  It's Gini coefficient is 0.3.  c)  Galaxy at $z = 0.74$ in the COSMOS field with a Gini coefficient of 0.3.  d)  Another galaxy from the COSMOS field with a Gini of 0.3.  This one at $z = 0.65$.}
\label{fig:ngc4790}
\end{center}
\end{figure*}

The strong correlation between concentration and M20 is due to the fact that they have similar definitions (see section 3.6).  As mentioned in section 3.6, however, M20 is much more sensitive to non-axisymmetric features which occur in many disks as well as in mergers.  This is why outliers lying above (higher M20) the M20-C relation begin to appear as one moves towards the low-C, high-M20 part of the plane.

Figure~\ref{fig:casgm-local} shows the distribution of our values of C, A, S, G and M20 compared to the one in the samples of \citet{Conselice03} and \citet{Lotz04}.  Because of the fact that we effectively use the flux within one petrosian radius in our estimate of the total flux (see section 3.1) as opposed to $1.5 \times r_{petro}$ as in \citet{Conselice03}, our values of C are lower by about 0.3.  From theoretical considerations of S\'{e}rsic profiles \citep{Sersic68}, we obtained a similar shift rather ubiquitously across all S\'{e}rsic indices going from 1 (exponential) to 4 (de Vaucouleurs).  Other than that constant shift, all distributions are otherwise fairly similar, except for the fact that \citet{Conselice03} and \citet{Lotz04} have a higher fraction of objects with bulge-dominated morphologies.  This, however, has no implications since their sample was selected by hand.

The asymmetry values in our sample are also slightly different ($\Delta A \sim 0.1$).  This is certainly at least partly a real effect as higher redshift galaxies tend to show more peculiarity \citep[e.g.][]{vandenBergh96}.  However, we did implement a different background asymmetry subtraction algorithm, which is specially written so that to avoid oversubtraction.  Our clumpiness(S) measurements are also systematically lower.  Such a decrease in S with redshift was predicted and calculated, however, by \citet{Conselice03}, and is a consequence of bright knots getting smeared out in images with lower resolution and lower signal-to-noise, the latter being a consequence of surface brightness dimming.  In addition, the higher the real clumpiness, the larger this effect is.  Our clumpiness values, thus rarely exceed $0.5$.

Our Gini values also tend to be lower, but we think this is also simply due to lower signal-to-noise.  Figure~\ref{fig:ngc4790}a shows the picture of NGC 4790, an Sd galaxy, taken from the Digitized Sky Survey.  Because it has a very uniform disk, it's Gini coefficient is low, 0.4 in this case, which is the low-end limit in \citet{Lotz04}'s sample, but which is not yet as low as many objects in our sample.  However, we found that by degrading its image to the level shown in figure~\ref{fig:ngc4790}b, we were able to reproduce an image with a Gini coefficient of 0.3, similar to the ones of figure~\ref{fig:ngc4790}c and~\ref{fig:ngc4790}d which represent two $G \sim 0.3$ galaxies from our sample.  On the other hand, we find the Gini coefficient of ellipticals to be a fairly robust to resolution and signal-to-noise effects, which means that image degradation only stretches the low-end of the {\em Gini}-distribution, and indeed, this is what we observe in figure~\ref{fig:casgm-local}.  Finally, our values of M20 correspond very well to those of \citet{Lotz04}.  The only difference being, as we mentioned earlier, their higher fraction of bulge-dominated objects.

\subsection{Physical Properties of $z \sim 0.7$ Galaxies}

\begin{figure*}[thbp]
\begin{center}
\includegraphics[scale=0.9]{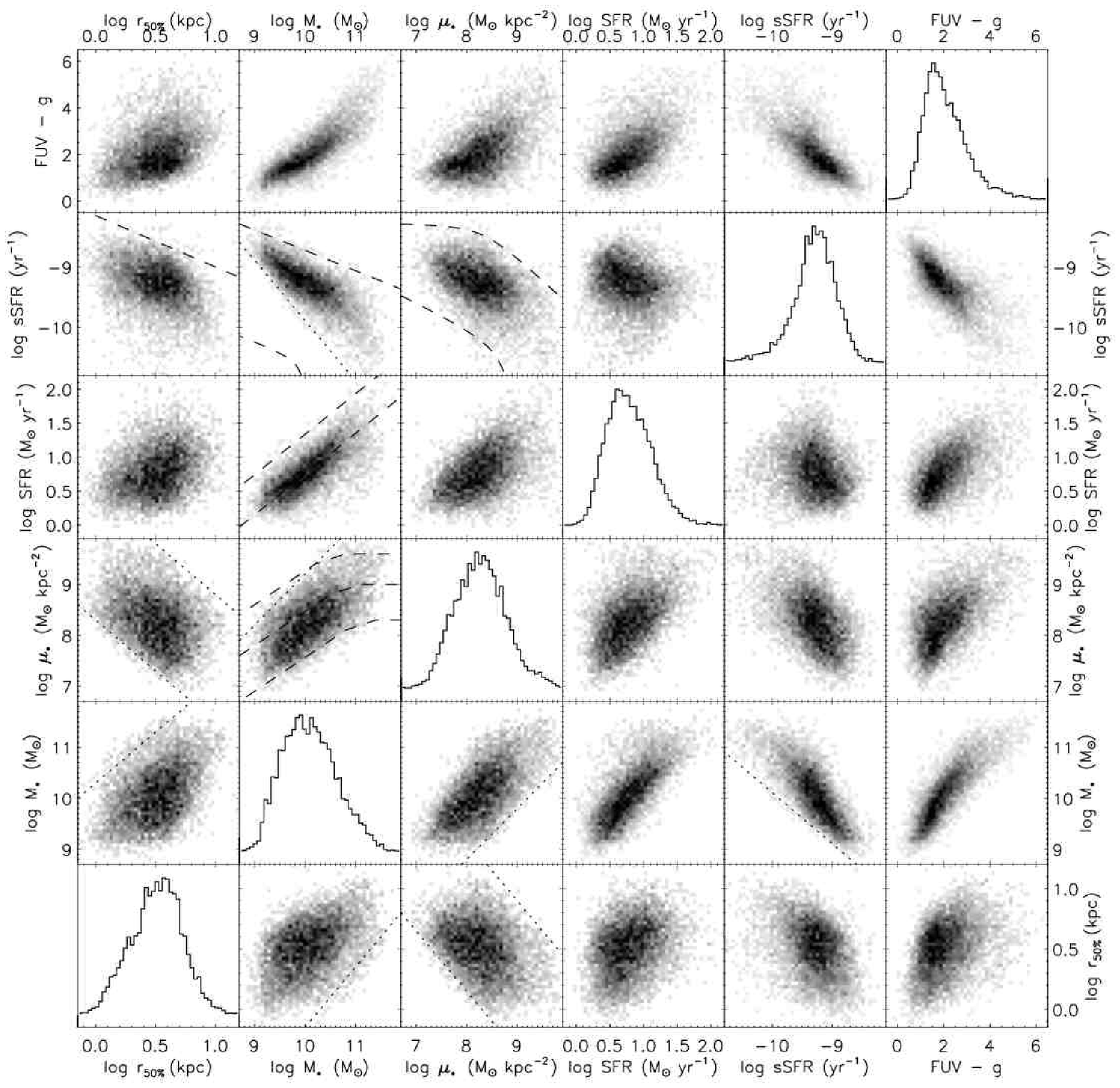}
\caption{Distribution of our UV-detected sample in various physical parameters.  Dotted lines represent in the $\log M_{*} - \log$ sSFR plane:  our detection limit of log SFR = 0.11; in the $\log \mu_{*} - \log r_{50\%}$ plane:  our detection limit of $\log \mbox{ M}_{*} = 9.1$ and the line of $\log \mbox{ M}_{*} = 11.6$ (which is the high-mass cutoff in our mass distribution); in the $\log \mu{*} - \log M_{*}$ plane:  our detection limit of $r_{50\%} = 1.0$ kpc;  in the $\log M_{*} - \log r_{50\%}$ plane:  the line of $\log \mu_{*} = 9.5$ (which represents the typical value for ellipticals).  Dashed lines represent in the log sSFR vs. $\log r_{50\%}$ plane:  the upper envelope of the \citet{Brinchmann04} local relation shifted up by 0.45;  in log sSFR vs. $\log M_{*}$:  the upper and lower envelopes of the \citet{Brinchmann04} local relation shifted up by 0.55;  in log sSFR vs. $\log \mu_{*}$:  the upper and lower envelopes of the \citet{Brinchmann04} local relation shifted up by 0.45; in log SFR vs $\log M_{*}$:  the median for blue galaxies as well as the upper envelope of the local relation of \citet{Brinchmann04} obtained from their raw fiber measurements (their figure 17), both shifted up by 0.35;  and in $\log \mu_{*}$ vs $\log M_{*}$:  the upper and lower envelopes as well as the median of the local relation from \citet{Kauffmann03b} (unadjusted).}
\label{fig:rmuss-withuv}
\end{center}
\end{figure*}

\begin{figure*}[thbp]
\begin{center}
\includegraphics[scale=0.9]{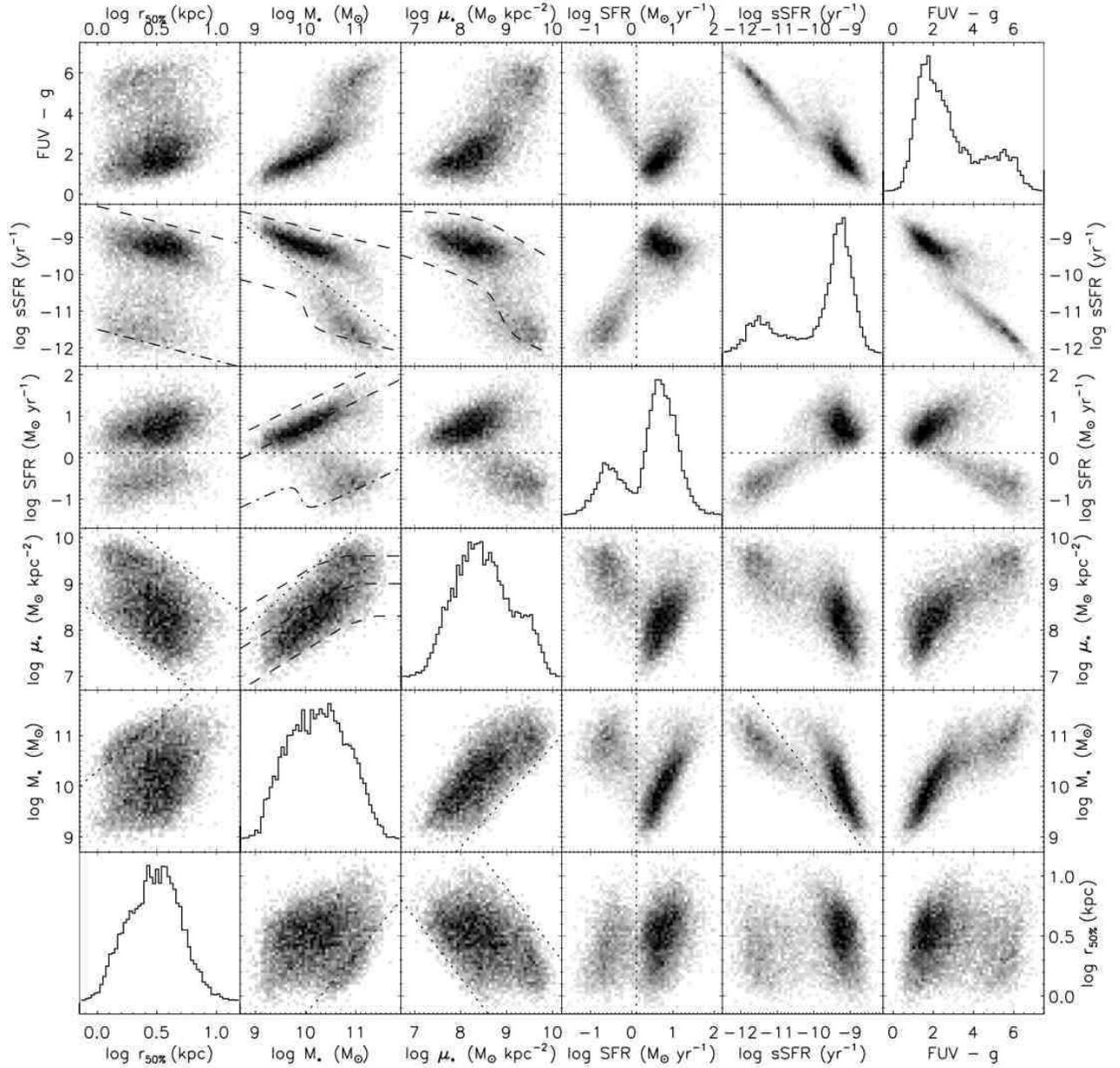}
\caption{Same as figure~\ref{fig:rmuss-withuv},  but for our full sample.  Star formation rates below our detection limit have been obtained through SED extrapolation and are thus prone to large uncertainties.  Dotted lines have been added in plots involving SFR that represent our detection limit of log SFR = 0.11.  Additional dashed-dotted lines are also shown that represent the (unadjusted) lower envelopes of the relations obtained by \citet{Brinchmann04} for local galaxies.  }
\label{fig:rmuss}
\end{center}
\end{figure*}

\begin{figure*}[thbp]
\begin{center}
\includegraphics[scale=0.9]{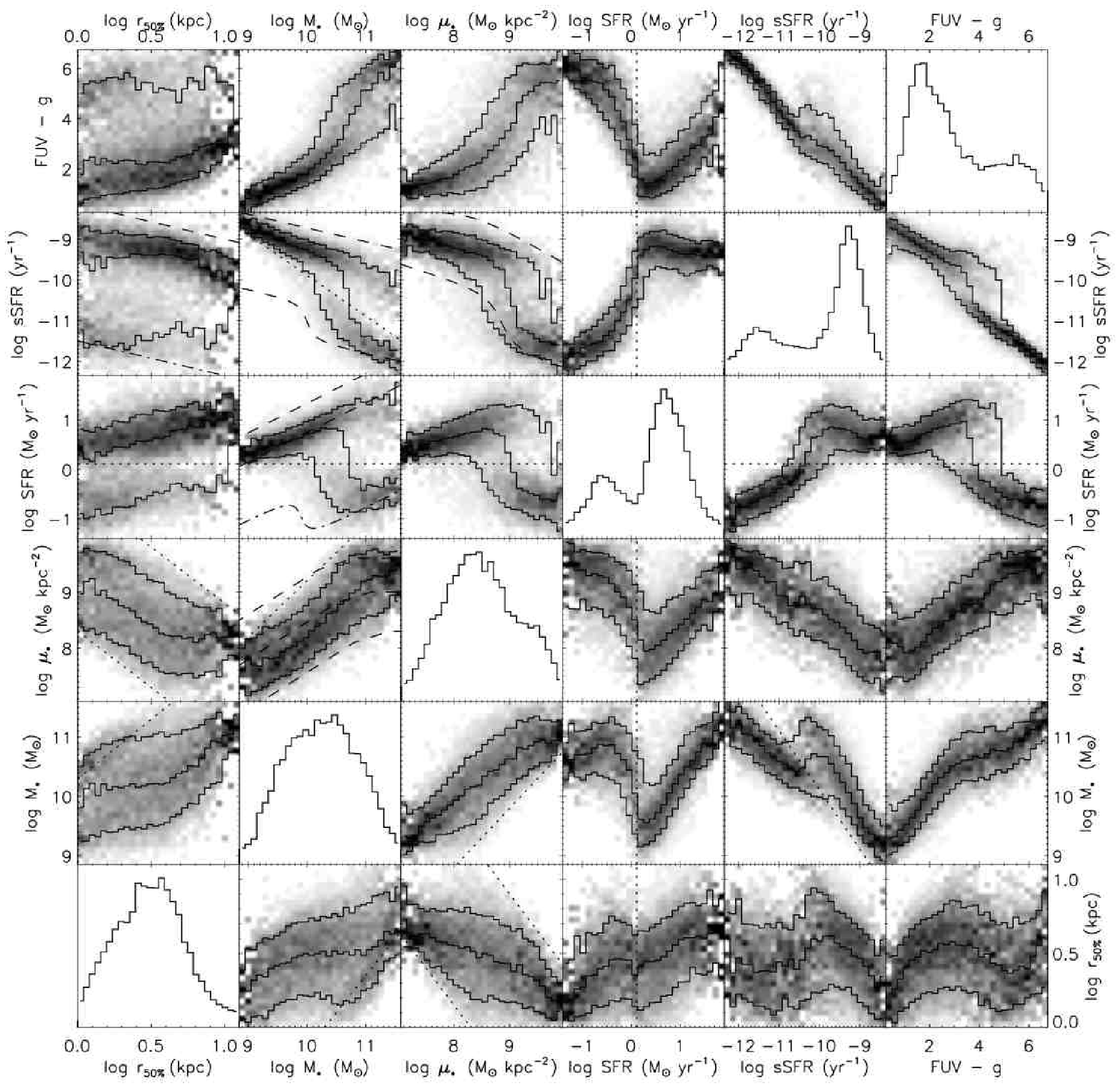}
\caption{Conditional plot of figure~\ref{fig:rmuss}, that is with each column being normalized separately.  The solid lines represent the 10, 50, and 90\% quantiles in every column.  We reiterate that star formation rates below our UV-detection limit have been obtained through SED extrapolation and are prone to large uncertainties.}
\label{fig:rmuss-quantiles}
\end{center}
\end{figure*}

Figures~\ref{fig:rmuss-withuv} and~\ref{fig:rmuss} represent the distribution of sizes, masses, surface mass densities, star formation rates, specific star formation rates and restframe FUV - $g$ color in our UV-detected and full samples respectively.  Size, mass and star formation rate measurements were performed as described in sections 3.1, 2.2 and 2.4 respectively.  Surface mass densities were derived using the following relation: 
\begin{equation}
\mu_{*} = \frac{0.5 M_{*}}{\pi r_{50\%}^{2}}
\end{equation}
where the factor of 0.5 accounts for the average effect of inclination.  We also plot, in figure~\ref{fig:rmuss-quantiles}, the conditional plot of figure~\ref{fig:rmuss}, that is a plot where each column has been normalized separately, i.e. the value in each bin has been divided by the total number of objects in that same range of the independent variable.  We also show in figure~\ref{fig:rmuss-quantiles}, for every column, the 10, 50 (or median) and 90\% quantiles.  These clearly trace out how the dependent variable varies specifically as a function of the independent variable as the effect of the number of objects present at a given value of the independent variable is removed by the normalization.  They also provide us with a sense for the spread in the relation.
 
\begin{figure}[hbtp]
\begin{center}
\includegraphics[width=3.4in]{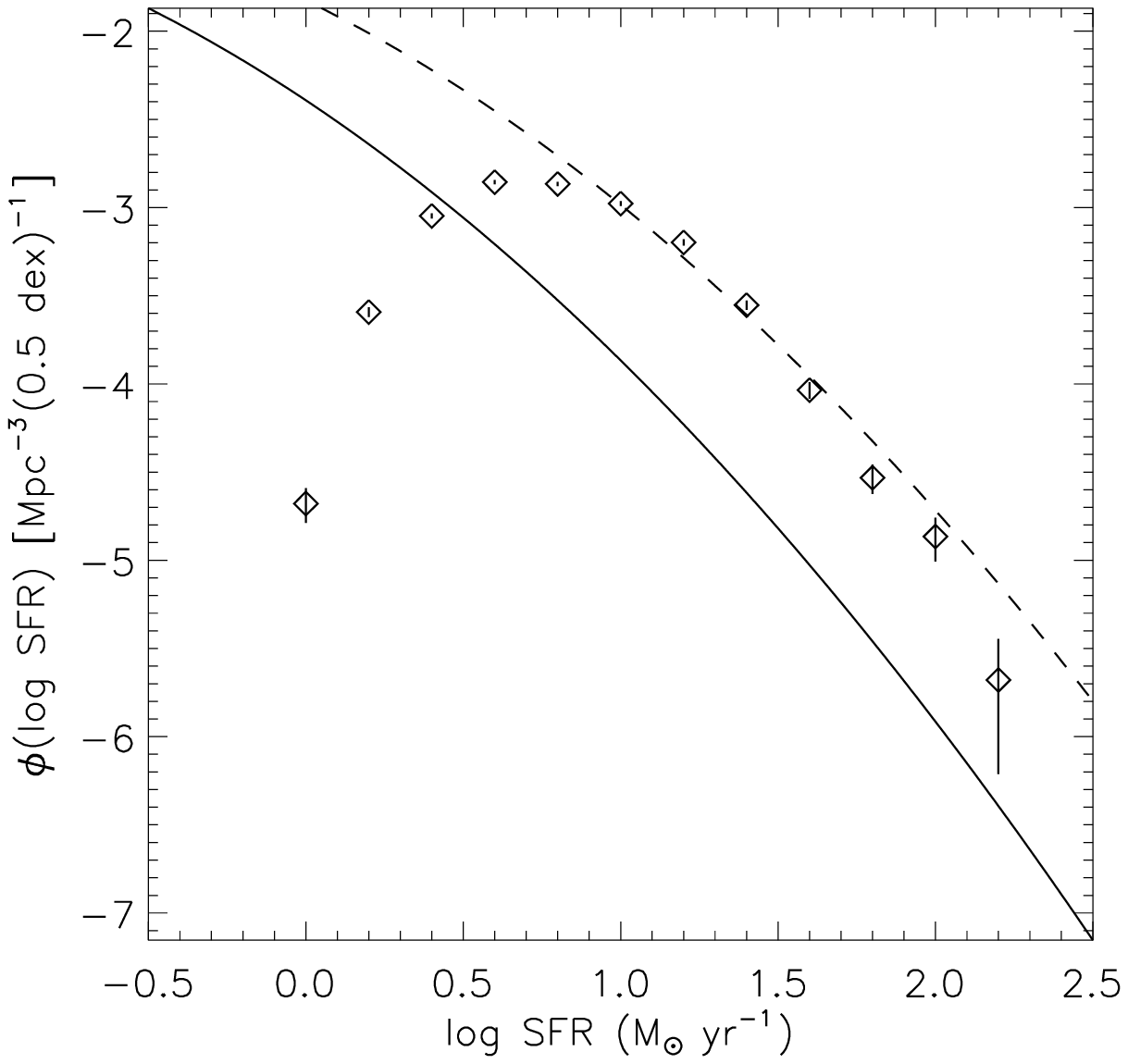}
\caption{Star formation rate distribution in our UV-detected sample compared to the local star formation rate function of \citet{Martin05b} (solid line) and to the same function shift by 0.55 in log SFR (dashed line).}
\label{fig:sfr_function}
\end{center}
\end{figure}

As our selection criteria are optimized for morphological analysis, our sample is only complete for large ($r_{50\%} \gtrsim 3$ kpc) and massive ($\log M_{*} \gtrsim 10$) galaxies.  We nevertheless observe the properties of galaxies in our sample to be, within our completeness limits, consistent with that of local galaxy samples in sizes, masses and surface mass densities.   They do, however, have higher star formation rates.  This increase in star formation rate also reflects itself in figure~\ref{fig:sfr_function},
 which shows our star formation rate distribution in comparison to the local star formation rate function of \citet{Martin05b}.  In fact, figure~\ref{fig:sfr_function} demonstrates that our distribution is shifted by a factor of about 3.5 (0.55 in log space) which corresponds to the overall increase in star formation rate density of the Universe between $z =0$ and $z = 0.7$ \citep{Schiminovich05}.  We demonstrate in the second half of this section that, by comparing to local samples, this shift is ubiquitous among the full range of star-forming galaxies.  As for the color distribution, it displays bimodality (figure~\ref{fig:rmuss}), similar to that observed at low redshifts \citep{Strateva01}, as well as very distinct red and blue sequences when plotted against mass \citep{Wyder06}, with the transition occurring at $\log M_{*} \approx 10.5 \mbox{ M}_{\odot}$.  When plotted against star formation rate, it shows that the range of possible SFR's widens as one progresses to redder colors with the population bifurcating into old quiescent galaxies on one side, and dust-enshrouded star-forming ones on the other.  Many of the relations in figures~\ref{fig:rmuss-withuv} to~\ref{fig:rmuss-quantiles} have been studied in the low-redshift population by \citet{Kauffmann03b} and \citet{Brinchmann04}.  We thus now turn to compare our results with theirs to investigate whether or how these relations have changed since redshift $z \sim 0.7$.

\citet{Kauffmann03b} observed $\log \mu_{*}$ to be proportional to $\log M_{*}$ for $\log M_{*} \lesssim 10.5$ followed by a flattening at higher masses.  The relation between $\log \mu_{*}$ and $\log M_{*}$ in our UV-detected sample (figure~\ref{fig:rmuss-withuv}) is consistent with theirs, but comparison of the \citet{Kauffmann03b} relation with our full sample (figures~\ref{fig:rmuss} and~\ref{fig:rmuss-quantiles}) is harder to reconcile.  Within our detection limits, our two relations are consistent at low-masses, but we fail to observe the break in the slope for higher-mass objects.  However, because of our incompleteness at small radii, we are missing the top part of the relation (where $r_{50\%} \lesssim 1.0$ kpc), which deprives us of an upper envelope for the linear part of the relation thus making it hard to conclude one way or the other.  A flattening at high masses would still be consistent with our results in the case that the upper envelope for galaxies at $z \sim 0.7$ be higher than in the \citet{Kauffmann03b} relation.

We also observe $r_{50\%}$ to behave, as a function of mass, in a way similar to that reported by \citet{Kauffmann03b} for local galaxies and \citet{Barden05} for samples with redshift up to $z \sim 1$, namely slowly increasing at low masses followed by an accelerated increase and truncation of small size galaxies at masses $\gtrsim 10^{10.5} \mbox{ M}_{\odot}$ with a typical spread of about an order of magnitude in $r_{50\%}$.  The median of our relation is slightly shifted towards larger radii compared to the trend found by \citet{Kauffmann03b} for local populations, but this is simply an artifact of our incompleteness at low radii.  Our distribution is close, though, to the one of \citet{Barden05} for similar redshifts.

Analogously to \citet{Brinchmann04}, we also find a correlation between log SFR and $\log M_{*}$ at low masses, with a break at $\log M_{*}/M_{\odot} \gtrsim 10$, where a bimodality begins to emerge (figure~\ref{fig:rmuss-quantiles}).  The \citet[][figure 17]{Brinchmann04} relation for fiber measurements of both SFR and $M_{*}$ augmented by 0.35 in log SFR is shown in dashed lines in figures~\ref{fig:rmuss-withuv},~\ref{fig:rmuss} and~\ref{fig:rmuss-quantiles} (median and upper envelope representing 0.02 conditional probability contour) and is the relation that fits best our star-forming galaxies (UV sample).  The dotted-dashed line shown in figures~\ref{fig:rmuss} and~\ref{fig:rmuss-quantiles} represents the unmodified lower contour and appears to form a better envelope of our sample.  It is thus possible that only the star-forming galaxies have higher star formation rates at $z \sim 0.7$, though the star formation rates below our GALEX detection limit do carry large uncertainties.

We do also observe the well known, both locally \citep{Brinchmann04,Perez-Gonzalez03} and at higher redshifts \citep{Feulner05,Brinchmann00,Cowie96}, relation between specific star formation rate and mass.  Figures~\ref{fig:rmuss} and~\ref{fig:rmuss-quantiles} again demonstrate a bimodality and spread in the relation at masses $M_{*}/M_{\odot} \gtrsim 10$.  We have seen that a shift of 0.35 in log SFR from the \citet{Brinchmann04} fiber measurements fits best our data.  They do calculate, however, an aperture correction (for both SFR and mass) and apply it to their study of specific star formation rate.  Comparing with the aperture corrected data, the shift necessary to bring the log sSFR - log M relation of \citet[][figure 24]{Brinchmann04} to fit ours is now 0.55 in log SFR, precisely the value representing the difference in $\rho_{SFR}$ of the Universe between $z \sim 0.7$ and present \citep{Schiminovich05}.  Our specific star formation rates are slightly higher, however, than the ones found by \citet{Feulner05} for the same redshift.  We, nevertheless, believe our results to be accurate as our distribution differs from theirs mostly at the high-mass end where we see a number of galaxies with significant ($\gtrsim 10 \mbox{ M}_{\odot} \mbox{ yr}^{-1}$) star formation rates, whereas their results show an upper envelope at SFR $\approx 5 \mbox{ M}_{\odot} \mbox{ yr}^{-1}$.  As an aside, since specific star formation rate is correlated with the ratio of present to average past SFR, the behavior we observe corroborates the evidence that more massive galaxies formed their stars, on average, earlier than lower-mass galaxies, also referred to as downsizing \citep{Cowie96}.

Our relation between specific star formation rate and mass surface density is also qualitatively similar to the local relation, with a transition at $\log \mu_{*} \gtrsim 9$ from star-forming to quiescent objects.  This suggests that star formation ceases beyond a certain stellar density, in agreement with \citet{Kauffmann03b} and \citet{Brinchmann04}.  More precise comparison further shows that our upper envelope for the distribution function is shifted up by 0.45 in log sSFR relative to \citet{Brinchmann04}'s as plotted in figures~\ref{fig:rmuss-withuv} to~\ref{fig:rmuss-quantiles}.  Similarly, the relation between specific star formation rate and size for our UV-detected sample is consistent with the local distribution upped by 0.45 in log sSFR (figure~\ref{fig:rmuss-withuv}).  Although in both relations the lower contour appears not to change from the local distributions, the large uncertainties in the star formation rate estimates below our NUV detection limit make it hard to speculate on the evolution of low star-forming galaxies.

Since our mass surface density and size distributions do not appear to evolve much from $z = 0.7$ to the present, it is safe to assume that the observed evolution in specific star formation rate simply reflects the evolution in star formation rate itself.  It thus appears that the relations between physical parameters of $z \sim 0.7$ galaxies change little relative to local samples but for an overall increase by a factor of $10^{0.35}$ to $10^{0.55}$ in their star formation rate.  As $10^{0.55}$ precisely represents the decline in star formation rate density in the Universe from $z = 0.7$ to $z=0$, this overall dimming with time appears to account for most of the star formation rate density evolution since that redshift, confirming the results of \citet{Wolf05}.  On the other hand, since our observed shift could be as low as $10^{0.35}$, our results can easily accomodate a mild number evolution in the form of a steepening of the faint-end slope in the FUV-luminosity function such as observed by \citet{Arnouts05}.

\subsection{Relation of Morphological Parameters to Physical Properties}

In this section we look at how morphological parameters relate to physical properties of galaxies.  The most important result we discover is the fact that all morphological parameters show a transition at masses of $10^{10.5} M_{\odot}$ and above, which we interpret as a shift from disk-dominated to bulge-dominated galaxies.  This transition mass is identical to the one observed by \citet{Kauffmann03b} for local galaxies and consistent with that observed by \citet{Bundy05} also in the redshift range $0.55 < z < 0.8$.  We also discuss in this section how incompleteness in the physical properties discussed in the previous section affects our distribution of morphological parameters.

\begin{figure}[hbtp]
\begin{center}
\includegraphics[width=3.4in]{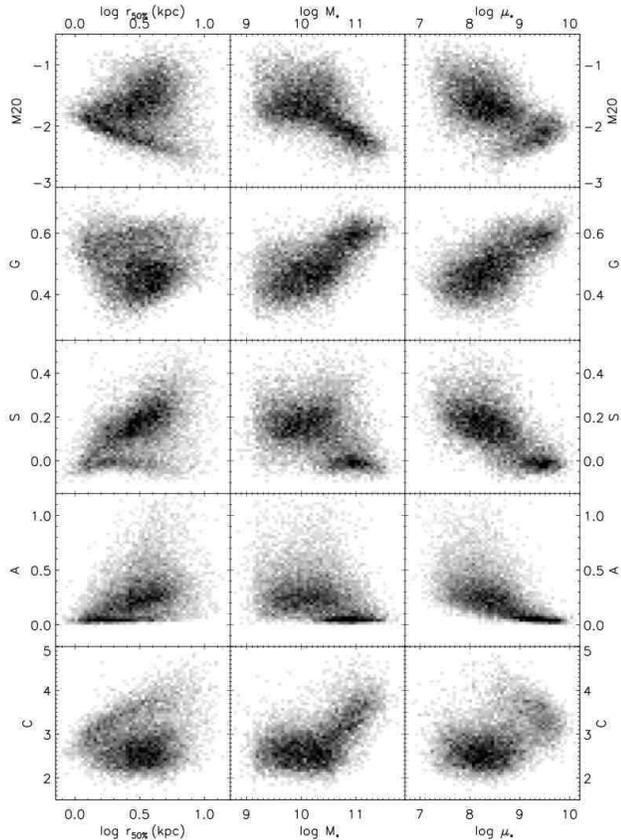}
\caption{Morphological parameters as a function of radius, mass and surface mass density for our full sample.}
\label{fig:casgm-vs-rmu}
\end{center}
\end{figure}

\begin{figure}[hbtp]
\begin{center}
\includegraphics[width=3.4in]{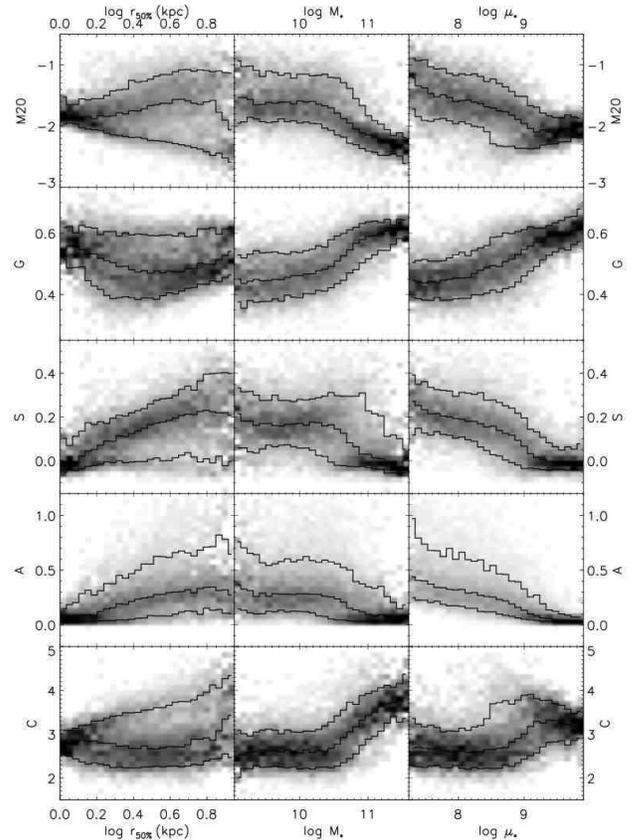}
\caption{Conditional plot of figure~\ref{fig:casgm-vs-rmu} with each column being normalized separately and lines representing the 10, 50, and 90\% quantiles in every column.}
\label{fig:casgm-vs-rmu-quantiles}
\end{center}
\end{figure}

\begin{figure*}[hbtp]
\begin{center}
\includegraphics[scale=0.9]{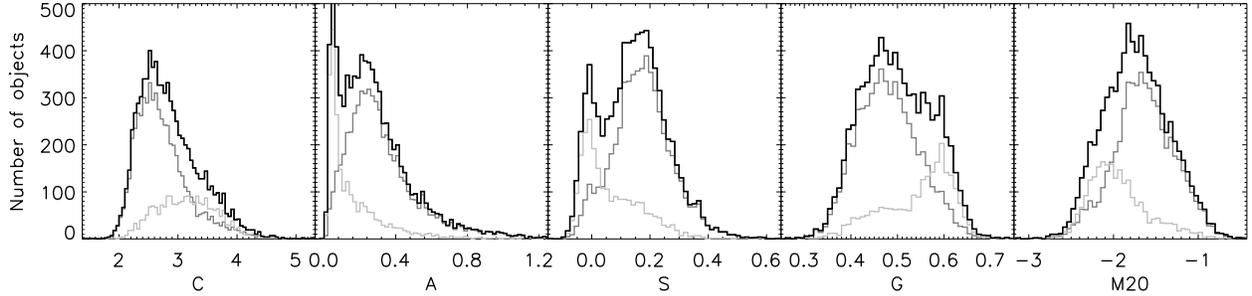}
\caption{Distribution of our sample in our five morphological parameters.  Black histograms show the distribution of the entire sample, while dark and light grey histograms respectively represent objects detected and objects not detected in the UV.}
\label{fig:casgm-histo}
\end{center}
\end{figure*}

Figures~\ref{fig:casgm-vs-rmu} and~\ref{fig:casgm-vs-rmu-quantiles}
 display each of our morphological parameters as a function of size, mass and surface mass density for our full sample, while figure~\ref{fig:casgm-histo} compares the distribution in morphology of galaxies from our UV-detected and non-UV-detected samples.  The large scatter in all the relations shown in figure~\ref{fig:casgm-vs-rmu} indicates that no one morphological parameter is a good indicator of the physical state of a galaxy.  Nevertheless, we do observe some broad correlations as well as a clear bimodality in some cases, both of which we discuss in this section.  We focus here on figures~\ref{fig:casgm-vs-rmu} to~\ref{fig:casgm-histo}, and postpone discussion of how star formation relates to morphology to the next section.

Concentration appears to show some kind of relation to most of our studied physical parameters, the most interesting one being that with stellar mass.  Figures~\ref{fig:casgm-vs-rmu} and~\ref{fig:casgm-vs-rmu-quantiles} indeed show how the distribution of galaxies in the $C - \log M_{*}$ plane displays a sharp transition at $\log M_{*} / M_{\odot} = 10.5$.  This is much the same behavior as seen locally by \citet{Kauffmann03b}, and indicates a shift from a disk-dominated population for which $C \approx 2.6$ independently of mass to a bulge-dominated population whose concentration rapidly increases with mass.  We find C to also behave in a similar way with $\log \mu_{*}$.  There appears, on the other hand, to be little correlation between concentration and size in our sample for the most part except for the fact that the point-spread function has a greater effect on the central profile of smaller galaxies, which causes the observed cutoff at progressively lower concentrations as we go to smaller $r_{50\%}$.

We have already discussed how the distributions of asymmetry and clumpiness show a bimodality with bulges and disks segregating at the low and high-end respectively.  This can be seen again in figure~\ref{fig:casgm-histo} as well as in most relations of A and S in figure~\ref{fig:casgm-vs-rmu}.  For example, although asymmetry and clumpiness do not seem to correlate with stellar mass in neither disk-dominated nor bulge-dominated galaxies, the two populations differentiate in their respective values of A and S,  bulges lying almost exlcusively in the range of  0 to 0.1 for A and -0.05 to 0.05 for S, and disks spanning almost the entire range of possible values, with typical values around $A = 0.25$ and $S = 0.17$.  We again see the transition from a disk-dominated population to a bulge-dominated one to take place at masses of around $10^{10.5} M_{\odot}$.  Although a bimodality can also be seen in the A or S vs. log $\mu_{*}$ planes (figure~\ref{fig:casgm-vs-rmu}), the two populations definitely follow a more linear relation with both A and S monotonically declining with $\log \mu_{*}$ (figure~\ref{fig:casgm-vs-rmu-quantiles}).  Asymmetry and clumpiness also appear to correlate with size in disk galaxies.  We demonstrate in the appendix how such a relation between clumpiness and size is expected in a resolution-limited sample.  As for asymmetry, it is unclear whether the observed trend is real or artificial.  Certainly, our brightness cut of $I < 23 \mbox{ mag}$ corresponds, at any given radius, to a surface brightness limit which progressively increases as we go to smaller radii.  As spheroids have higher surface brightnesses than disks, this selection effect can skew the observed population at a given size, which in turn influences the median asymmetry, and contributes to the observed trend.

Beyond the bimodality, we also note the crest extending to high values of A in all the plots of A versus physical parameters.  This crest is interesting since it represents galaxies that depart from the standard disk model by displaying asymmetric features, whether intrinsic or due to interactions.  It lies straight above the normal disk population, implying that these galaxies do not have physical properties different from those of regular disks (though other properties, that we do not study, such as IR-luminosity or dynamics, would probably differ).

The Gini coefficient is the only morphological parameter to correlate with mass throughout our entire mass range, even though the steepest increase still occurs at masses between $10^{10.5}$ and $10^{11} M_{\odot}$.  On the other hand, {\em Gini} is mostly insensitive to the size of the galaxy, and hence correlates equally well with surface mass density.  Nonetheless, we observe a lack of low-{\em Gini} small objects which is due to surface brightness incompleteness combined with the fact that the Gini coefficient systematically increases at low resolution where objects approach a point-like psf profile.  We also observe an increase, with size, of the low-end cutoff of the {\em Gini} distribution at larger radii ($\log r_{50\%} > 0.4$).  This increase appears to be real, and signifies that larger disks are allowed less uniformity in their light distribution, implying they are either more bulgy, more clumpy or have, in general, more structures, such as spiral arms.  Often, all three apply.

Moving to M20, as shown in figure~\ref{fig:casgm}, it is largely anti-correlated with concentration.  It hence displays many of the same trends as concentration, only mirrored.  For example, M20, just like concentration, appears to be independent of mass for $\log M_{*} \lesssim 10.5$, but strongly (anti-)correlated with it at higher masses.  A notable difference, however, is that, as seen in figure~\ref{fig:casgm-histo}, star-forming galaxies occupy two thirds of the possible range of values of M20 (-2.0 to -0.5 relative to a full range going from -2.75 to -0.5), whereas they occupy only half of the possible range of values of concentration (from 2.0 to 3.25 relative to a range of 2.0 to 4.5).  This is because M20 is more sensitive than C to bright features far from the center.  This reflects in the relation between M20 and $\log r_{50\%}$, which shows a bifurcation.  As discussed previously for concentration, the downward trend in bulge-dominated galaxies is simply due to our resolution limit.  On the other hand, just like in the case of asymmetry, it is unclear whether the upward trend in disk-dominated galaxies is real or also a consequence of our selection criteria.  For one, the smallest diffuse disks do not make our magnitude cut, which eliminates high-M20 objects at low-$r_{50\%}$, and then M20 and clumpiness are somewhat correlated so that resolution effects could also help create the observed trend.  Nevertheless, though it might be accentuated by selection criteria at small radii, it is likely that the relation be real as it holds at $\log r_{50\%} \ge 0.5$, our completeness limit, which suggests that larger disks are more likely to have bright regions in their outskirts, and that smaller galaxies have the bulk of their light distribution more centralized.

In summary, we have seen in this section how all morphological parameters show, as a function of stellar mass, a transition at $M_{*} \sim 10^{10.5} M_{\odot}$ which we interpret as a transition from disk-dominated to bulge-dominated objects.  We have also shown how the Gini coefficient is the only morphological parameter to correlate with stellar mass throughout our mass range.

\subsection{Morphology of Star-forming Galaxies}

\subsubsection{Star Formation Rate as Related to Morphological Parameters}

\begin{figure*}[hbtp]
\begin{center}
\includegraphics[scale=0.9]{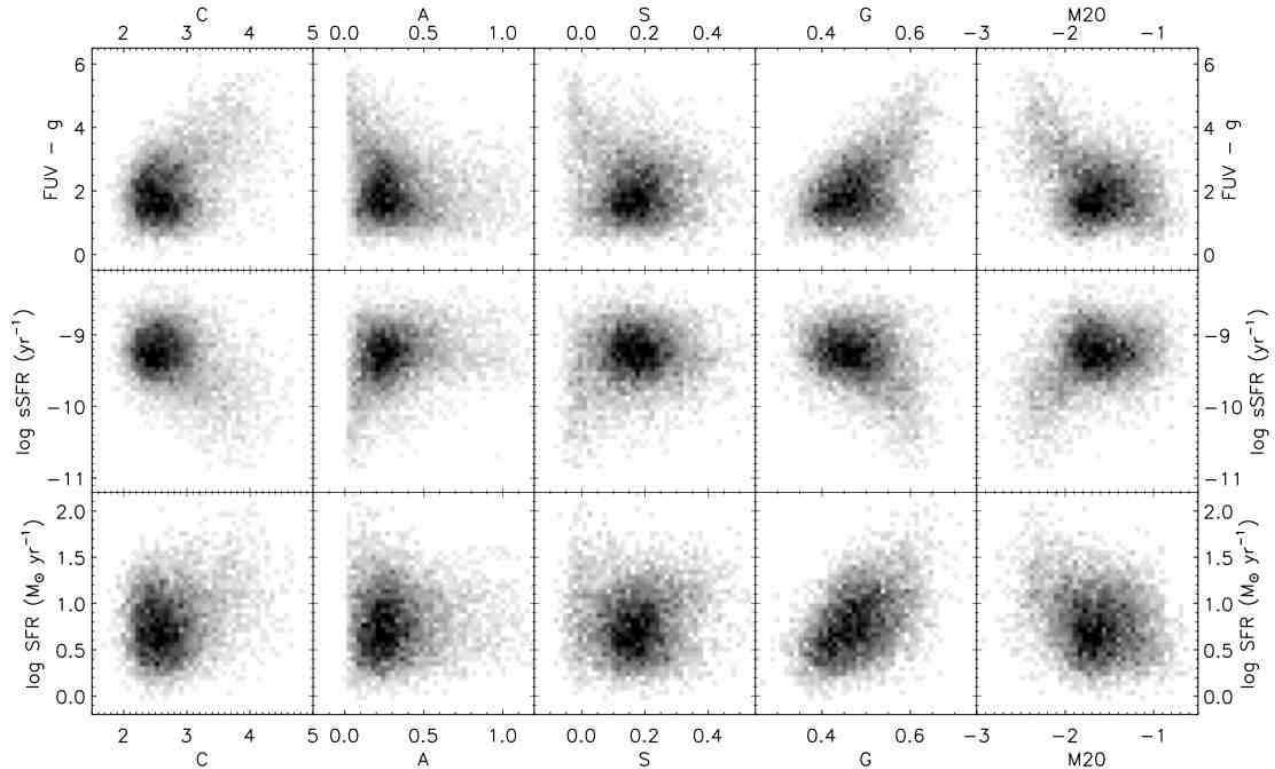}
\caption{Absolute and specific star formation rates as well as FUV - $g$ color of our UV-detected sample as a function of morphological parameters.}
\label{fig:sfr-vs-casgm}
\end{center}
\end{figure*}

\begin{figure*}[hbtp]
\begin{center}
\includegraphics[scale=0.9]{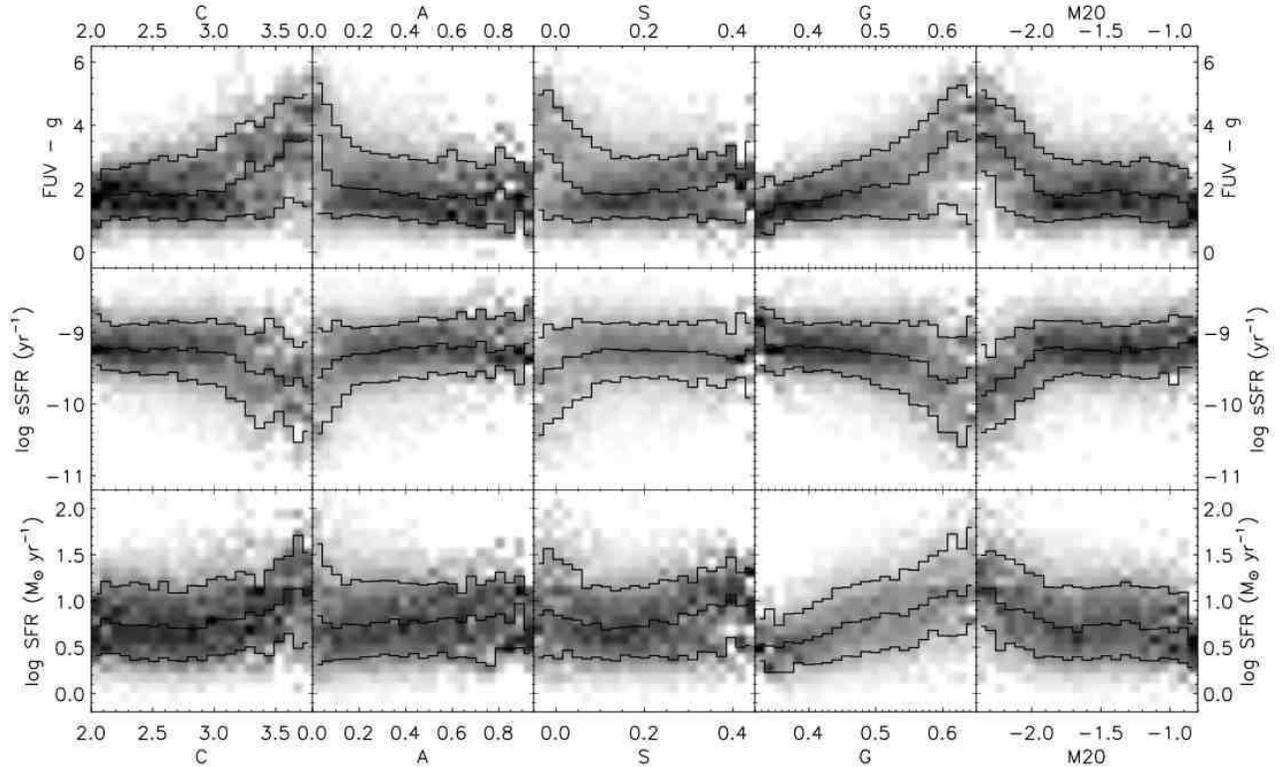}
\caption{Conditional plot (i.e. with each column separately normalized) of the absolute star formation rate, specific star formation rate and restframe FUV - $g$ color of our UV-detected sample as a function of morphological parameters.   The lines represent the 10, 50 and 90\% quantiles in every column.}
\label{fig:sfr-vs-casgm-quantiles}
\end{center}
\end{figure*}

\begin{figure*}[hbtp]
\begin{center}
\includegraphics[scale=0.9]{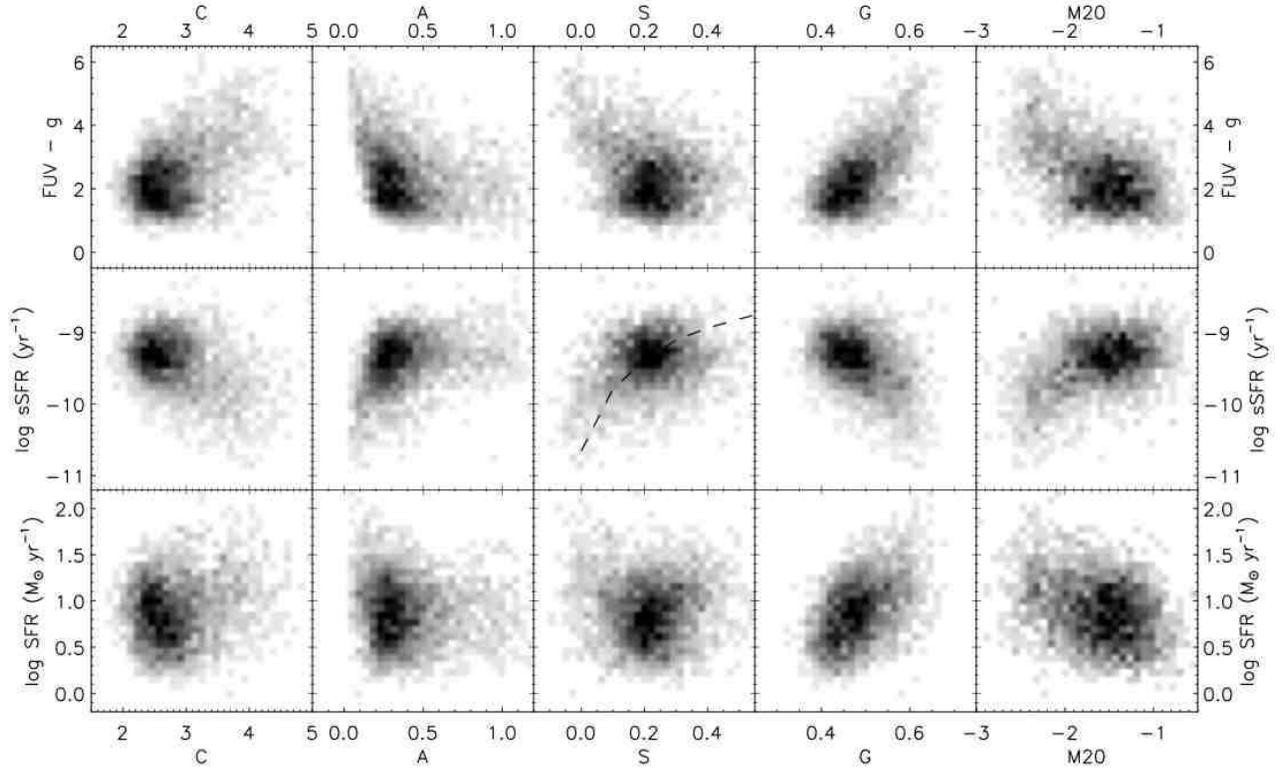}
\caption{Absolute star formation rate, specific star formation rate and restframe FUV - $g$ color of our UV-detected size-complete sample as a function of morphological parameters.  The dashed line in sSFR vs. S represent the relation of \citet{Conselice03} normalized to $z = 0.7$ and to our distribution in S.}
\label{fig:sfr-vs-casgm-complete}
\end{center}
\end{figure*}

\begin{figure*}[hbtp]
\begin{center}
\includegraphics[scale=0.9]{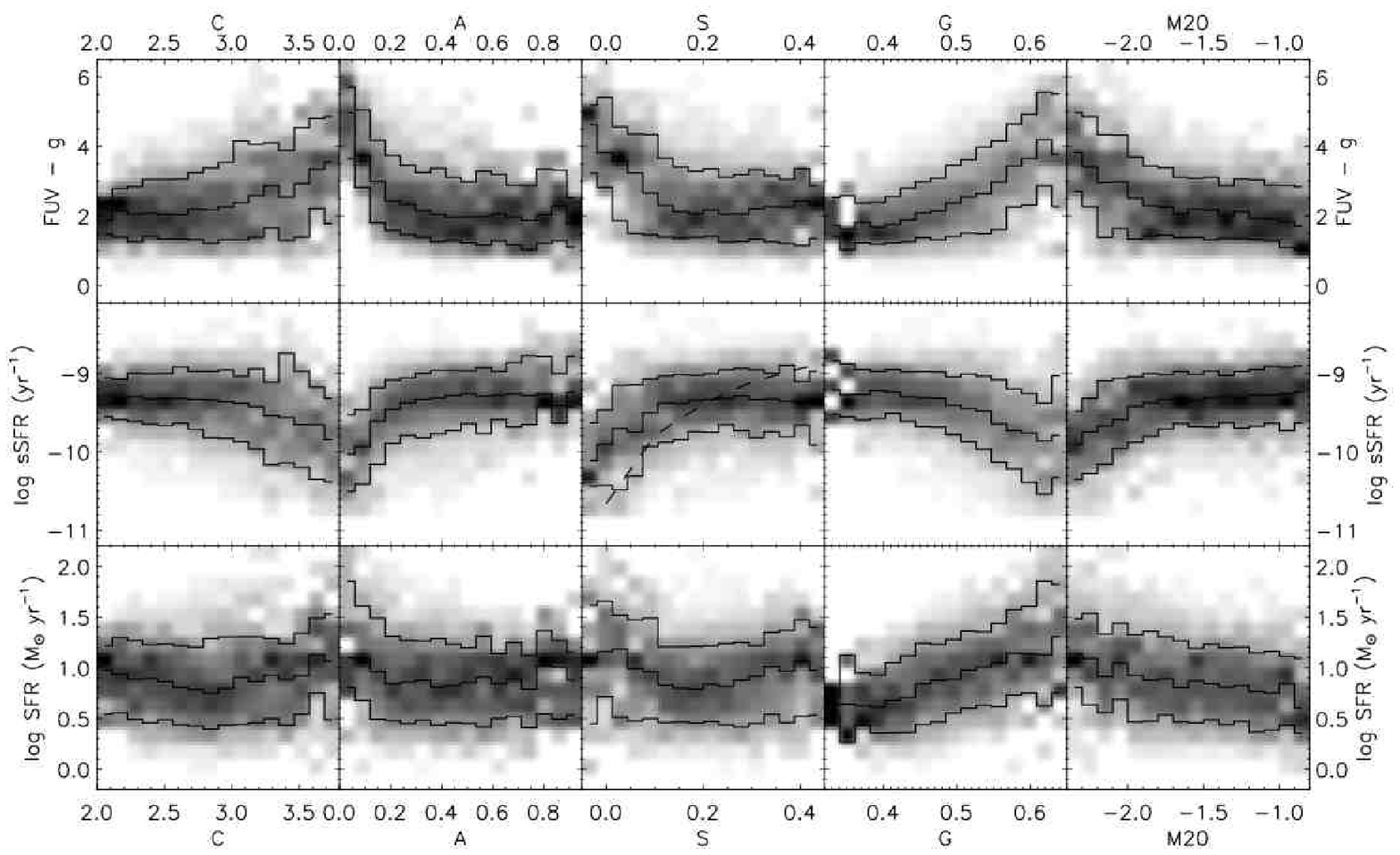}
\caption{Conditional plot (i.e. with each column separately normalized) of color, absolute and specific star formation rates of our UV-detected and size-complete sample as a function of morphological parameters.   The solid lines represent the 10, 50 and 90\% quantiles in every column.  The dashed line in sSFR vs. S represent the relation of \citet{Conselice03} normalized to $z = 0.7$ and to our distribution in S.}
\label{fig:sfr-vs-casgm-complete-quantiles}
\end{center}
\end{figure*}

The relation between color, star formation rate and specific star formation rate, and morphological parameters is shown in figures~\ref{fig:sfr-vs-casgm} and~\ref{fig:sfr-vs-casgm-quantiles} for galaxies detected in the UV.  As expected, disk galaxies are blue and become, on average, redder as the bulge becomes more prominent, that is as we move towards bulge-like morphologies.  Nevertheless, blue galaxies seem to exist for all values of our morphological parameters, including a population of blue compact galaxies.  Since many morphological parameters suffer from incompleteness at low radii, we also plotted in figures~\ref{fig:sfr-vs-casgm-complete} and~\ref{fig:sfr-vs-casgm-complete-quantiles} the same relations but for galaxies with $\log r_{50\%} > 0.55$ (or $r_{50\%} > 3.55$ kpc) only, and still taken from the UV-detected sample.  This cut represents our size completeness limit.  Relations of color to morphological parameters are stronger in this complete sample, but their qualitative behavior remains the same.  One notable difference is the absence of blue compact galaxies, due to the simple fact that they do not make the size cut.

From figures~\ref{fig:sfr-vs-casgm-quantiles} and~\ref{fig:sfr-vs-casgm-complete-quantiles}, it is clear that, among our five morphological parameters, the strongest relation color possesses is that with the Gini coefficient.  Color is also strongly correlated to asymmetry as well as to clumpiness for  $A \lesssim 0.2$ and $S \lesssim 0.12$.  However, it then quickly stabilizes and changes very little for values of $A \gtrsim 0.4$ and $S \gtrsim 0.2$.  These strong correlations at low values of A or S respectively confirm similar results of \citet{Conselice00} and \citet{Conselice03}.  The upper ranges of asymmetry and clumpiness, however, have not been probed before.  We find, here, that those highly asymmetric or very clumpy galaxies possess the same blue colors as their more standard disk counterparts.  Similarly, we find no significant difference, as discussed below,  in the star formation rate or in the specific star formation rate of high-A or high-S galaxies compared to normal disks.  We discuss our interpretation of the relation between color and morphology further in section 5.

Interestingly, the parameter with which star formation rate correlates the best is the Gini coefficient. Figure~\ref{fig:sfr-vs-casgm-quantiles} indeed shows that the logarithm of the star formation rate is a linear function of G (with log SFR $\varpropto 2.3 \times G$), although the relation has a dispersion of as much as 1 to 2 orders of magnitude (at the 80\% confidence level) from the lower to the upper end of the range in G.  In general, quiescent early-type galaxies have higher values of G than most of their star-forming counterparts, and actually dominate the high-G population as shown in figure~\ref{fig:casgm-histo}.   Star-forming galaxies that do, however, attain equally high values of G, seem to typically carry star formation rates $\gtrsim 10 \mbox{ M}_{\odot} \mbox{ yr}^{-1}$, which suggests that the strongest starbursts have very luminous components that stand out from the rest of the galaxy.  From inspection, we notice that these are often central components, possibly in the process of being formed.  We thus interpret this result as a signature that episodes of strong star formation can be often linked to bulge growth, growth that sometimes turns galaxies into early-type objects.  We, again, address this subject in more detail in section 5.

Otherwise, the star formation rates for our UV-detected galaxies, do not show a significant correlation with any other one of our morphological parameters, including asymmetry and clumpiness (though a slow decline of SFR with M20 is visible in our size-complete sample).  Given that all of our morphological parameters are normalized quantities, one might expect specific star formation rate, which is also a normalized quantity, to show a stronger correlation with morphological parameters, and in particular with clumpiness which was shown to correlate with the $H\alpha$ equivalent width by \citet{Conselice03}.  This is, however, hardly the case.  Specific star formation rate does show lower values for bulge-like morphologies and a rapid rise towards disk-like morphologies, but it then quickly tappers off so that among disks, the sSFR  becomes rather independent of morphology.  For ends of comparison, we have also plotted \citet{Conselice03}'s relation between sSFR (converted from EW($H\alpha$) and renormalized to redshift $z \sim 0.7$ by a factor of $10^{0.35}$) and clumpiness, in figures~\ref{fig:sfr-vs-casgm-complete} and~\ref{fig:sfr-vs-casgm-complete-quantiles}.  This shows that our two results are not inconsistent, only that there exists a larger variety of galaxies, especially at $z \sim 0.7$, than was first considered by \citet{Conselice03}.  In particular, these include large spirals which, although harboring substantial star formation, tend to have lower specific star formation rates because of their large masses (cf. the sSFR-mass relation in figure~\ref{fig:rmuss}).  Their spiral structure and star formation regions would, nevertheless, confer them high values of clumpiness.  On the contrary, lighter, more compact and more flocculent late-type disks, would typically have higher specific star formation rates than the typical grand-design spiral, though often lower values of clumpiness.  This diversity in the galaxy population adds scatter, so that it becomes difficult to infer specific star formation rates from morphology alone.  We further discuss the relation between specific star formation rate and clumpiness in the appendix.

\subsubsection{Morphology, Color and Specific Star Formation Rate}

In the last two sections we have investigated relations between morphological and physical parameters, and although we discovered some interesting trends, inference of physical properties of galaxies from any one morphological parameter is uncertain at best, because of the important scatter in all of the existing correlations.  We now examine whether the situation can be improved upon by considering two morphological parameters simultaneously.  We have already demonstrated in section 4.1 how this can help separate spirals and ellipticals.  We now take one step further, and study color, specific star formation rate and FUV attenuation as a function of multi-parametric morphology.

\begin{figure*}[hbtp]
\begin{center}
\includegraphics[scale=0.9]{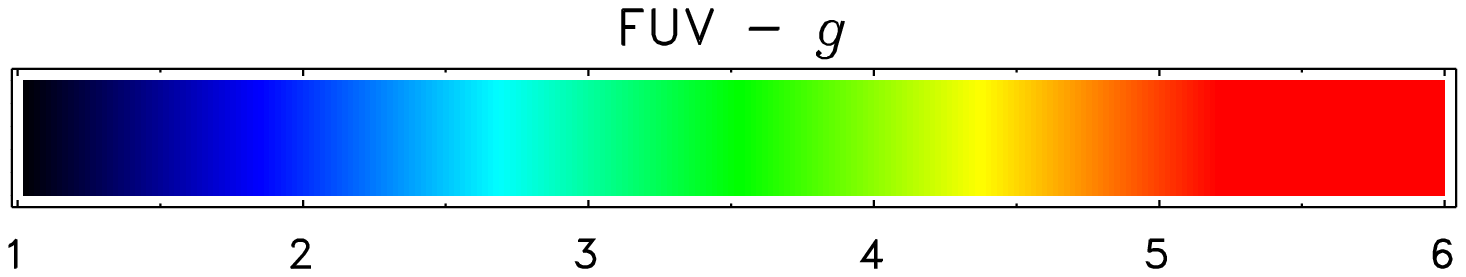}
\includegraphics[scale=0.9]{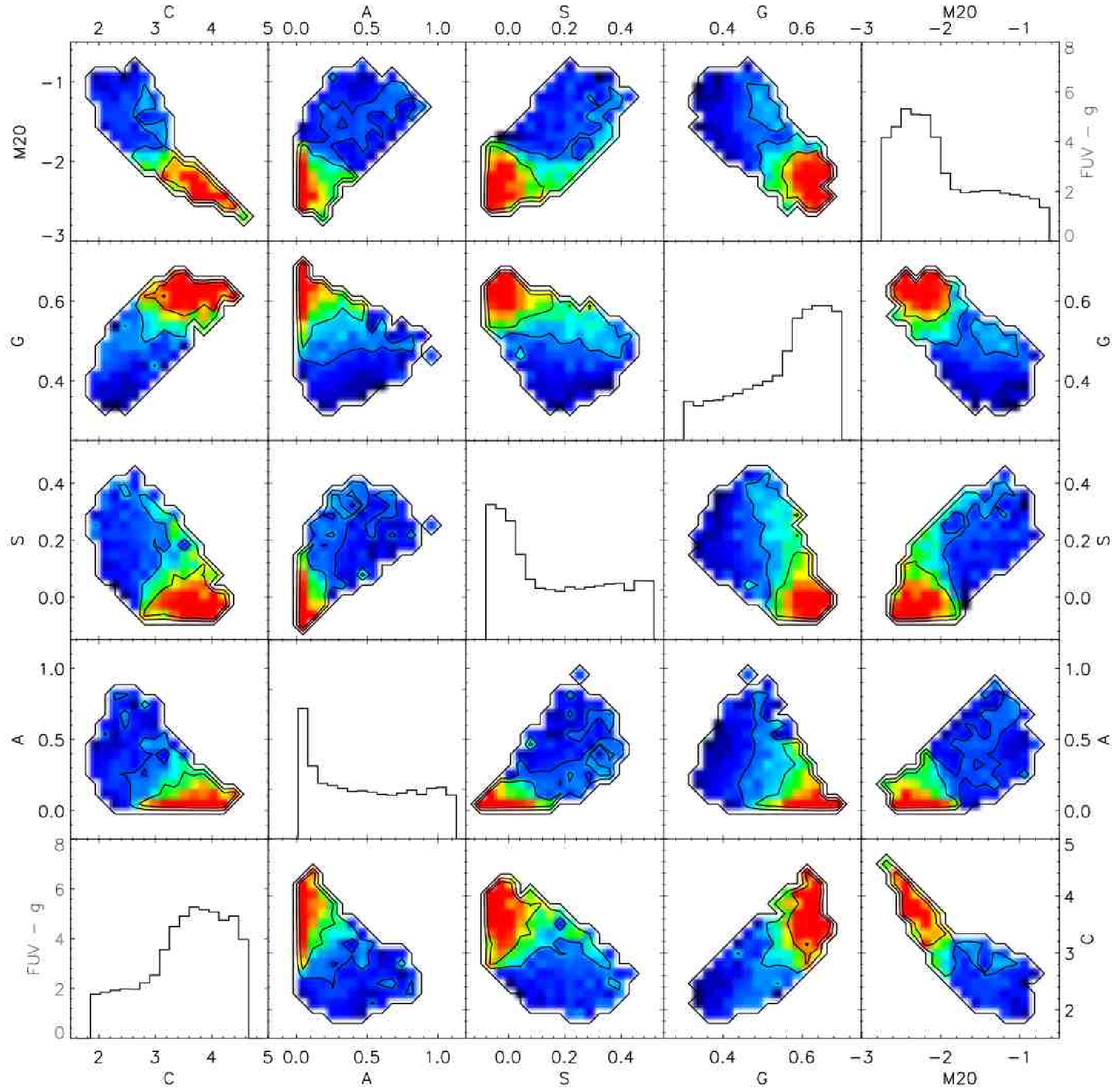}
\caption{\small Restframe FUV - $g$ color as a function of each as well as of each combination of two morphological parameters for our full sample.  The value in each color bin as well as in each histogram bin was obtained by taking the median FUV - $g$ color of all the objects present in that bin.  Bins with less than 10 objects were removed.  The color scale for histograms is indicated by the thin font top-most y-axis on the right side of the figure as well as the bottom-most y-axis on the left side of the figure.}
\label{fig:color}
\end{center}
\end{figure*}

\begin{figure*}[hbtp]
\begin{center}
\includegraphics[scale=0.9]{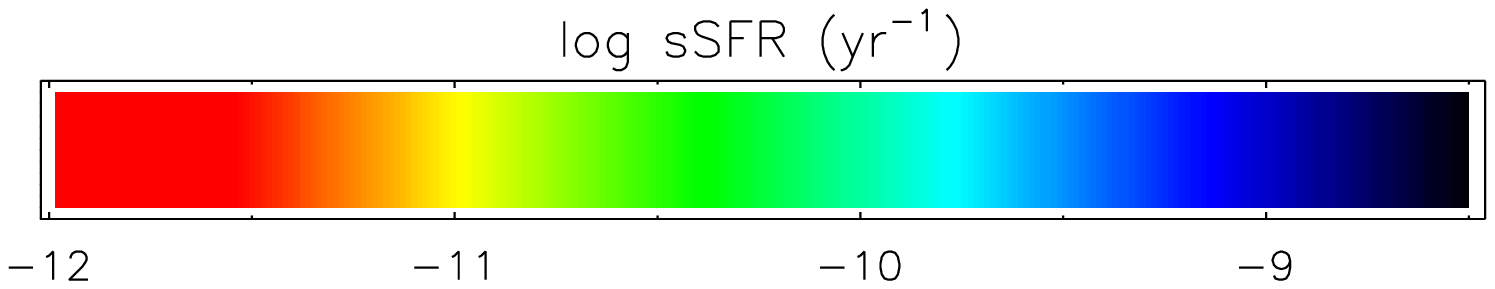}
\includegraphics[scale=0.9]{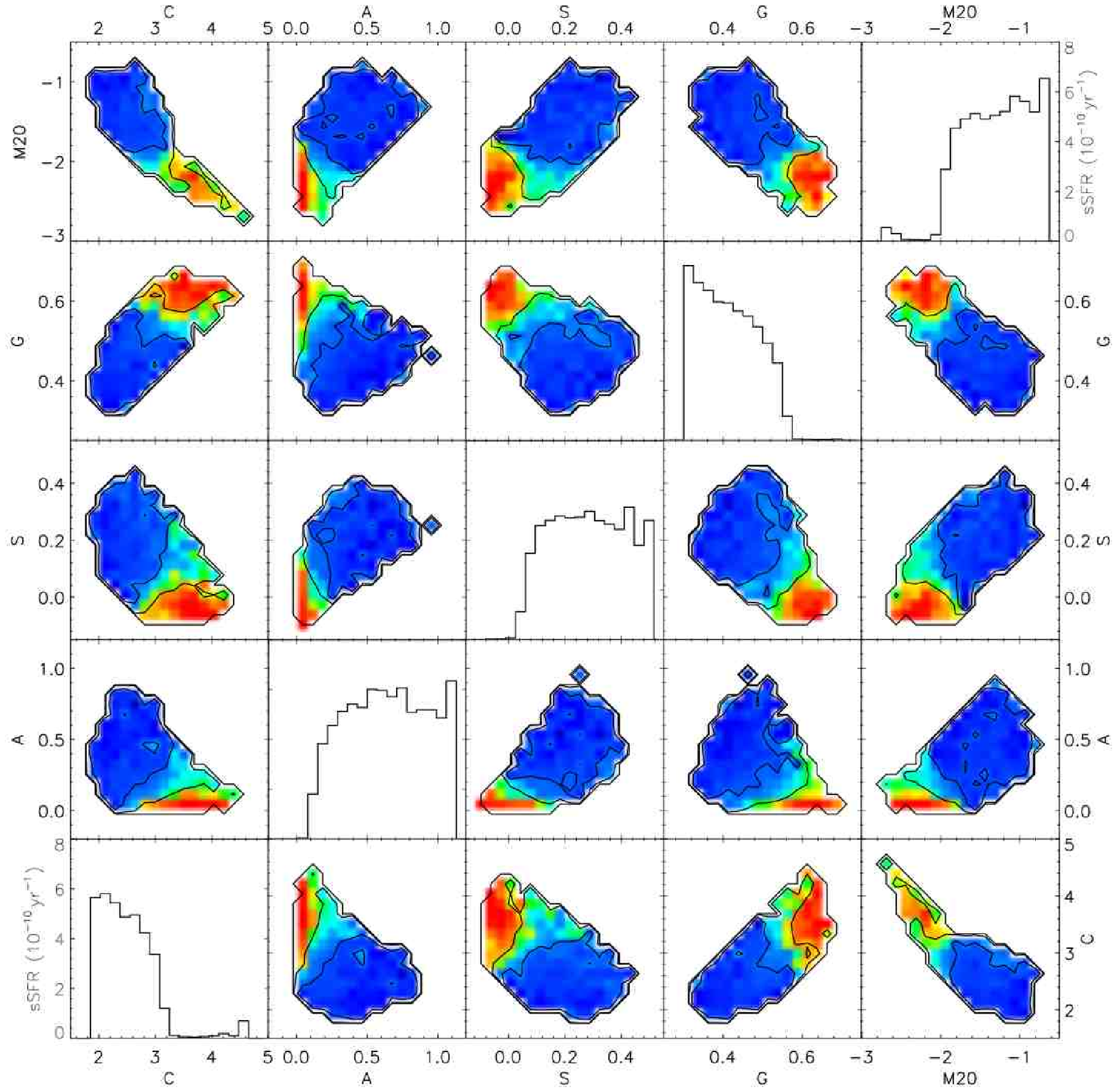}
\caption{Same as figure~\ref{fig:color} but weighted by specific star formation rate.}
\label{fig:ssfr}
\end{center}
\end{figure*}

\begin{figure*}[hbtp]
\begin{center}
\includegraphics[scale=0.9]{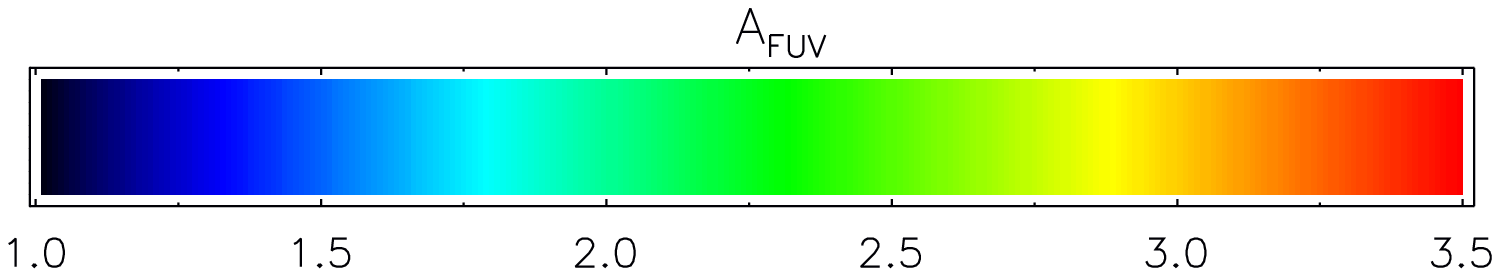}
\includegraphics[scale=0.9]{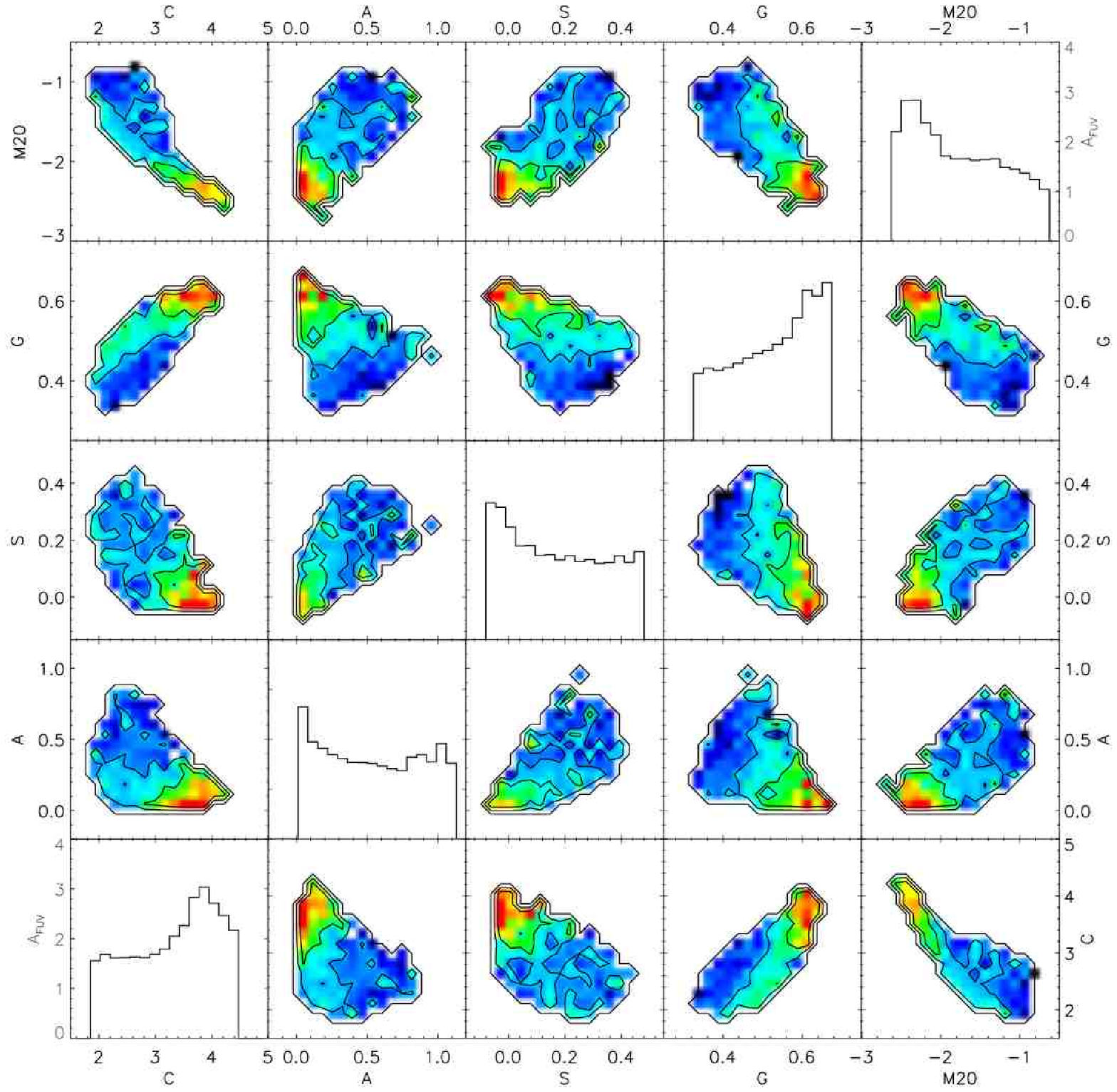}
\caption{\small FUV attenuation as a function of morphological parameters for UV-detected objects.  Redder represents more attenuation while bluer represents less dust correction.  The color displayed corresponds to the median value in that bin.  Bins with less than 10 objects were removed.  Histograms similarly represent median $A_{FUV}$ values across bins of each of our single morphological parameters, and are plotted on the scale indicated by the thin font top-most y-axis on the right side of the figure and the bottom-most y-axis on the left side of the figure.}
\label{fig:afuv}
\end{center}
\end{figure*}

Our approach has been to color code our morphology-morphology graph (figure~\ref{fig:casgm}) with the median of the studied quantity (FUV - $g$ color, sSFR or $A_{FUV}$) in a number of bins across the distribution.  The results are shown in figures~\ref{fig:color} through~\ref{fig:afuv}.  On the diagonal, we show the median distribution of FUV - $g$ color, sSFR and $A_{FUV}$ successively as a function every one morphological parameter.  We note that bins with fewer than 10 objects were discarded.

\begin{figure*}[htbp]
\begin{center}
\includegraphics[scale=0.9]{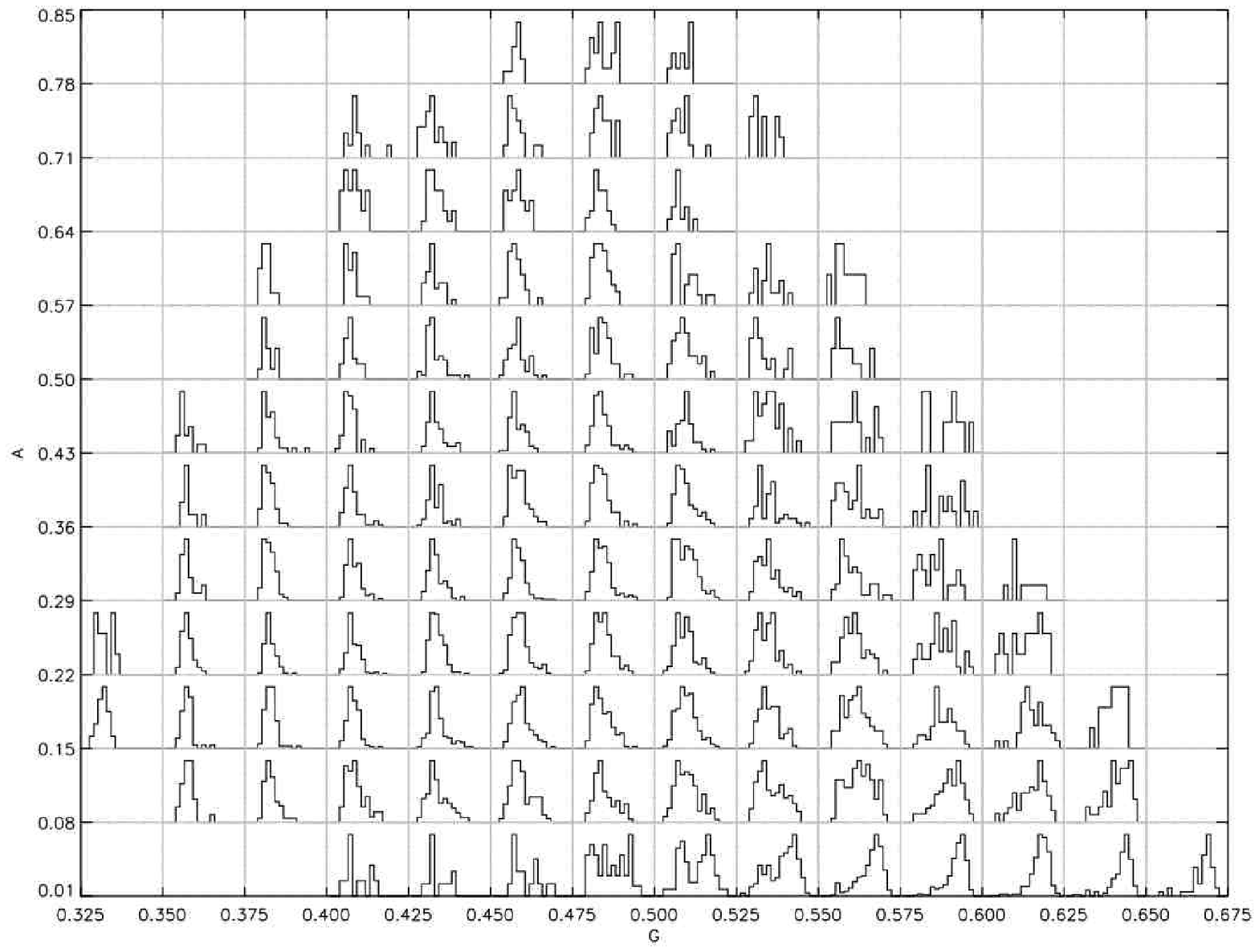}
\caption{Restframe FUV - $g$ color distribution of our full sample in bins of the A-G plane.  Every histogram spans a color range of $-1 <$ FUV - $g < 8$.}
\label{fig:gs_colorhists}
\end{center}
\end{figure*}

Figure~\ref{fig:color} demonstrates how, in the main galaxy population, the color and morphology bimodalities strongly overlap.  By comparing figure~\ref{fig:color} to figure~\ref{fig:casgm} one can see indeed, that the regions of morphological space where disk-dominated galaxies lie are blue and the ones populated by bulge-dominated ones are red.  Although this relation between morphology and color has been known for a long time \citep[e.g.][]{deVaucouleurs61}, we hereby demonstrate the strength of that correspondence in the particular case of FUV - $g$, as well as in the sharpness of the transition between the two populations.  Indeed, in most graphs of figure~\ref{fig:color}, the median color jumps from FUV - $g \approx 2$ (blue) to FUV - $g \approx 5$ (red) over only 2 or 3 morphological bins.  It is otherwise very little correlated with morphology within the two distinct populations themselves, and the color is thus very uniform over both regions of disk (blue) and bulge-dominated galaxies (red).  We note, however, that almost all of our bins, including the bulgy morphologies, do contain at least some blue galaxies.  In the bulge-dominated bins, these are the compact blue galaxies mentioned earlier.  Although very interesting, they are far less numerous and the majority of galaxies follow the pattern described above.  As a final point, we mention, that the converse, which would be bins of disk-dominated morphologies containing red galaxies, is not seen (figure~\ref{fig:gs_colorhists}).

Although most graphs in figure~\ref{fig:color} show quite a sharp transition, color seems to correlate more smoothly with the {\em Gini} coefficient than with any other morphological parameters, so that graphs that include G, but especially the A-G and S-G planes, show wider transition regions.  In the S-C and S-M20 planes, the region of concentrated galaxies that possess high clumpiness also shows intermediate median colors.   By inspecting the distribution of colors in those transition bins (as an example, we show the color distribution in bins of the A-G plane in figure~\ref{fig:gs_colorhists}),
 we discover that it is very wide, with all colors from FUV - $g$ = 1 to FUV - $g$ = 6 being represented more or less equally.  This indeed results into a typical median around 3.5, i.e. green.  As we move to bluer colors, we see that we begin to progressively lose the reddest objects, until we are left with only blue galaxies ($-0.5 <$ FUV - $g < 3$) in the dark blue bins.  In the red bins, on the other hand, red galaxies dominate over the other ones, though as just discussed, some blue galaxies might still be present.

\begin{figure}[htbp]
\begin{center}
\includegraphics[width=3.4in]{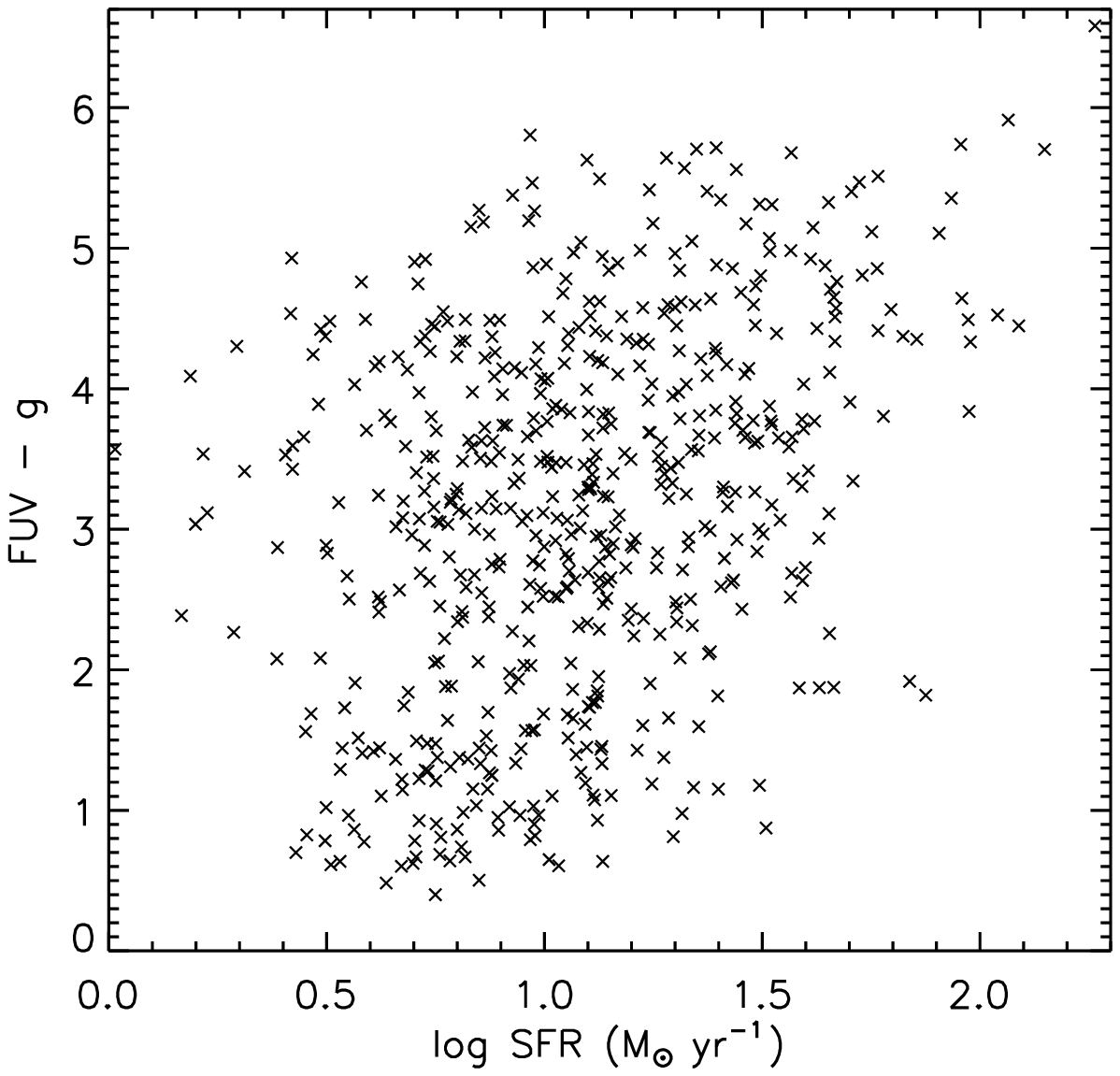}
\caption{FUV$-g$ color versus log SFR for UV-detected galaxies with a Gini coefficient of $G>0.57$.}
\label{fig:bulgy}
\end{center}
\end{figure}

The graph of specific star formation rate as a function of morphology (figure~\ref{fig:ssfr}) is qualitatively similar to the color-morphology one (figure~\ref{fig:color}).  However, its interpretation is a little different since the sSFR distribution is substantially more bimodal than the color distribution (figure~\ref{fig:rmuss}).  In that context, the median is more indicative of the number ratio between the two populations rather than the typical specific star formation rate of a galaxy in that bin.  The mode of the sSFR distribution of the quiescent population is situated, in our poor estimate, at around $\log \mbox{ sSFR} = -11.5$ (figure~\ref{fig:rmuss}), value that is represented by red in figure~\ref{fig:ssfr}.  This means that bins that include only quiescent galaxies are going to be red.  Orange bins, on the other hand, are likely already to contain a few star-forming objects, yellow bins even more, and so on.  The mid-point, where there is about an equal share of star-forming and quiescent galaxies, lies at $\log \mbox{ sSFR} \sim -10.4$:  green.  Once one reaches dark blue bins ($\log \mbox{ sSFR} \gtrsim -9.3$), only star-forming galaxies remain.  By looking at the various plots in figure~\ref{fig:ssfr}, it thus becomes apparent that most regions of morphology space, including the bulge-dominated areas, do contain at least some fraction of star-forming galaxies.  Some of the star-forming galaxies with bulge-like morphologies are the blue compact objects, but many of them are, instead, red and dusty. This is demonstrated in figure~\ref{fig:bulgy}
 where we show the FUV$-g$ color of galaxies possessing a Gini coefficient of $G > 0.57$, as a function of their star formation rate.  It also shows that the red and dusty galaxies are the ones that have the highest intrinsic star formation rates in our sample, though because they also tend to be massive, their sSFR is still typically lower than that of less massive galaxies (figure~\ref{fig:rmuss}).

To confirm the presence of dusty star-forming galaxies in bulge-dominated regions of morphology space, we plotted in figure~\ref{fig:afuv} the relation of $A_{FUV}$ to morphological parameters for our UV-detected sample.  All graphs indeed show a rise of $A_{FUV}$ in those regions.  These trends seem to indicate that the amount of attenuation could be mostly due to geometrical effects:  compact galaxies (high C and G, and low A, S and M20 galaxies) having the highest extinction and galaxies with extended star-forming disks or interacting objects with disturbed morphologies (high M20 or high A objects) possessing lower values of $A_{FUV}$.   However, as the regions of high extinction seem to coincide with those where spheroids are located, it is possible that they be contaminated by a number of bulge-dominated objects mimicking the effect of dust with older populations and thus modifying the observed relation between attenuation and morphology.  From inspection, we find that 77 out of 189 or $\sim 40\%$ of the UV-detected galaxies with $FUV - NUV > 1$ (which translates to $A_{FUV} > 4.12$) in our sample have early-type morphologies; the remaining 60\% show dusty-looking disks, though often also accompanied by an important central component.  Despite the fact that this is potentially substantial contamination, we do not believe this effect to significantly alter the trends in figure~\ref{fig:afuv} since dusty objects often possess larger UV-slopes than old populations anyway.  In order to confirm the robustness of our observations, we tried to identify and remove elliptical galaxies present in our UV-detected sample by, first, selecting them through various cuts in the space spanned by our morphological parameters, and then through visual inspection, but we always arrived at results very similar to those of figure~\ref{fig:afuv}.

Dusty starbursts are often luminous in infrared light.  \citet{Lotz04} have shown that the most extreme of these IR-luminous galaxies, ULIRGs, typically have high R-band values of M20, as they often carry signatures of interactions, but also high values of G, which is indicative of the fact that their optical light primarily comes from one or a few regions of concentrated luminosity (such as a nuclear starburst).  In our sample, the upper-right ridge of our M20-G plane shows higher FUV attenuation (figure~\ref{fig:afuv}) suggesting that galaxies in that part of the plane tend to be dustier.  Because of dust, galaxies that carry such morphologies also display redder colors than typical blue-sequence galaxies (figure~\ref{fig:color}) even though they possess similar specific star formation rates (figure~\ref{fig:ssfr}).  Our results thus appear to be consistent.  We also observe that many strongly star-forming galaxies lie in the same regions of morphology space as bulges.  These most likely represent more advanced stages of merging in which the resulting galaxy is finalizing its formation process.  Alternatively, these could also have grown from secular instabilities.  We explore the nature of starbursting galaxies in more detail in section 5.2.

It thus appears that the regions describing spheroidal galaxies (high-C, high-G and low-A, low-S and low-M20) are actually populated by a variety of objects:  dusty red starbursts, blue compact galaxies, quiescent elliptical galaxies, as well as UV-detected galaxies that, from visual inspection, also appear to be ellipticals.  We consider three possible origins for the presence of UV-light in elliptical galaxies:  residual star formation \citep[e.g.][]{Teplitz06,Yi05,Salim05,Stanford04,Menanteau01}, FUV-upturn from evolved hot horizontal-branch stars \citep{O'Connell99,Brown00,Rich05,Boselli05}, or weak AGNs.  We immediately rule out the FUV-upturn since it had been shown to fade rapidly with redshift reaching colors of $ FUV - V = 7$ at redshifts of $z \gtrsim 0.3$ \citep[][and references therein]{Lee05}, implying that we could only detect the FUV-upturn in galaxies with $m_{I} \lesssim 18.5$ which is brighter than any galaxy in our sample.  We cannot rule out any of the remaining two options, but we do see signs, in the asymmetry and/or clumpiness residual images, of faint underlying disks in about 50\% of our UV-detected spheroids with $FUV - NUV > 1$, suggesting that residual star formation does play an important role.  Although these galaxies appear to populate the lower half of star formation rates among UV-detected high-{\em Gini} objects (the upper half being mostly populated by dusty nuclear starbursts), their dust attenuation (and hence their SFR) could still be overestimated, particularly for galaxies with significant 500 to 800 Myr old populations, since such populations can carry steep UV-slopes \citep{Bruzual03}.

\subsubsection{Stellar Mass-limited Sample}

\begin{figure*}[hbtp]
\begin{center}
\includegraphics[scale=0.9]{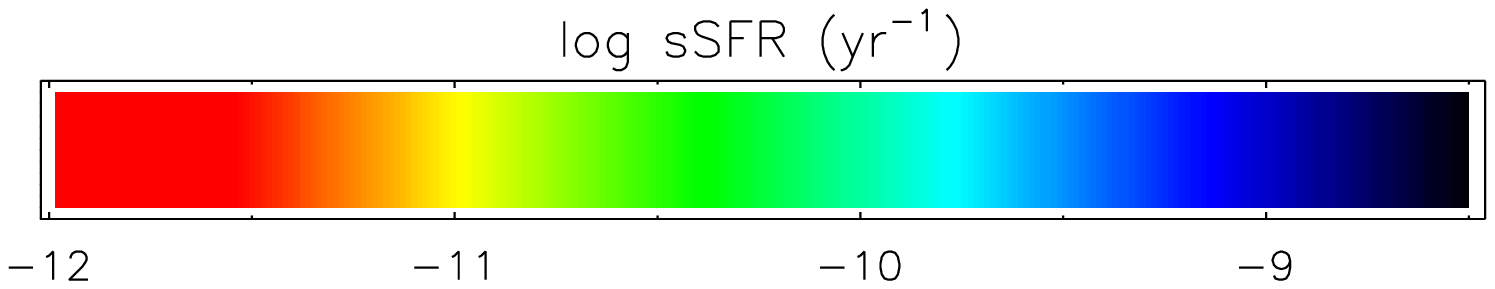}
\includegraphics[scale=0.9]{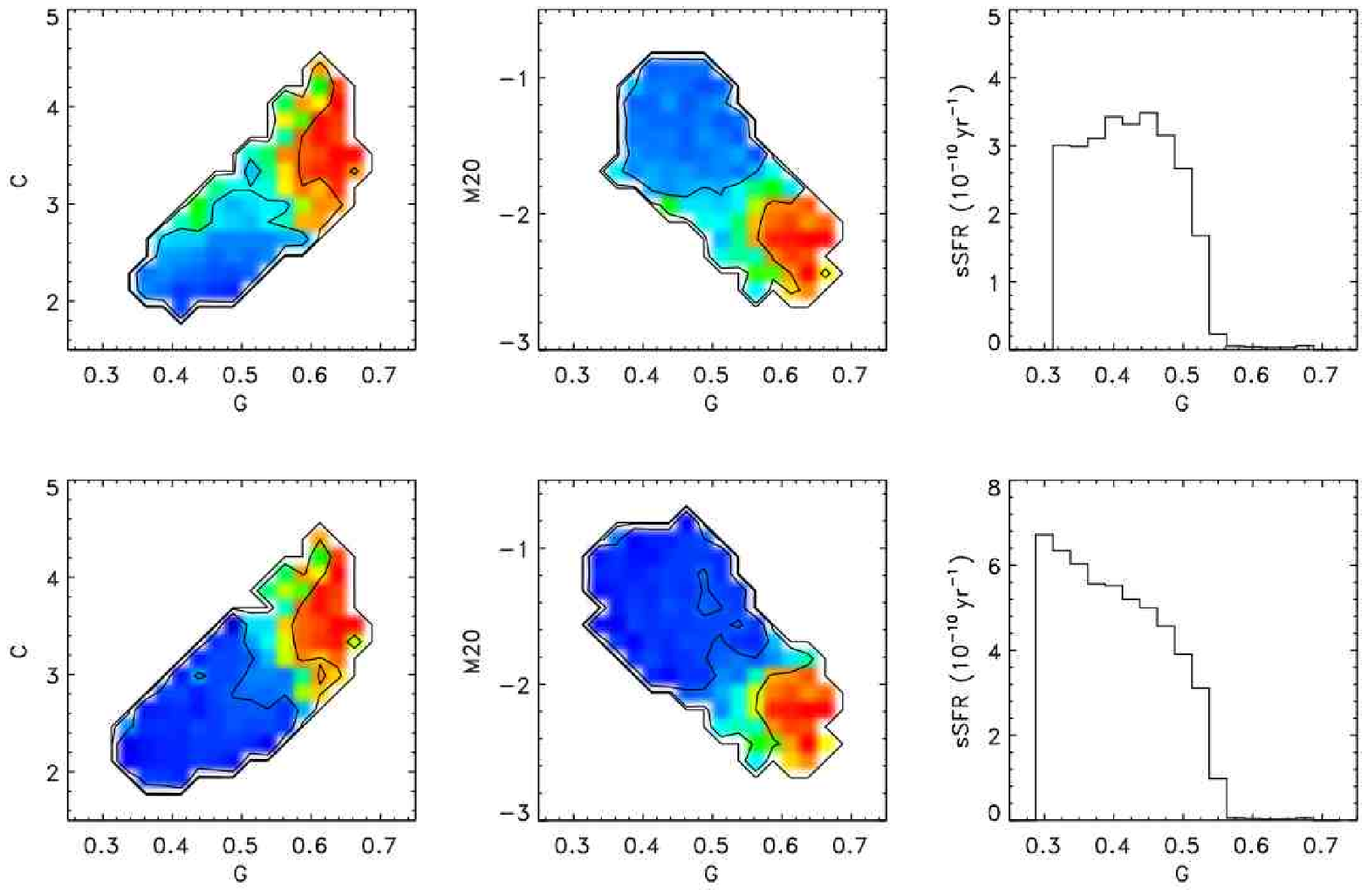}
\caption{Median specific star formation rate as a function of C and G (first column), M20 and G (second column), and G only (third column), for galaxies with $\log M_{*} \ge 10$ (top row) as compared to the same relations in our full sample (bottom row).  Note that the bottom row plots are identical to their figure~\ref{fig:ssfr} counterparts.  We show them again here to facilitate comparison with our mass-limited sample.  Also note the different axis scales on the two histograms.}
\label{fig:ssfr_masslimited}
\end{center}
\end{figure*}

As many studies nowadays utilize stellar mass-limited samples, we investigated how a mass cut would affect our results, focusing on figure~\ref{fig:ssfr}, the graph of specific star formation rate as a function of morphology.   We chose a stellar mass cut of $\log M_{*} \ge 10$, which is roughly our mass completeness limit.  Only in two instances do we observe significant differences with figure~\ref{fig:ssfr}: in C vs G and in M20 vs G.  The two plots for this mass-limited sample are shown in figure~\ref{fig:ssfr_masslimited} along with the histogram of the median sSFR as a function of G.  We also show in figure~\ref{fig:ssfr_masslimited} the same plots again for our full sample in order to facilitate comparison.  We observe that, in our mass-limited sample, the median sSFR flattens for values of $G \lesssim 0.5$, whereas it keeps rising at low values of G in our full sample.  The shape of the contours indicate that this difference originates primarily from the regions of the plots where $C \approx 3$ or $\mbox{M20} \approx -2$ and $0.4 < G < 0.55$.  
We think that such a difference in behavior at intermediate to low values of G can be simply explained by the fact that lower mass objects have lower signal-to-noise, and that galaxies with low signal-to-noise tend to have lower Gini coefficients (cf. \S4.1).  Lower mass galaxies thus tend to populate the upper-left half of the C-G plane, or the lower-left part of the M20-G plane, in higher numbers compared to the lower-right and upper-right halves respectively.  Because they have, on average, higher specific star formation rates than more massive galaxies (figure~\ref{fig:rmuss}), they tend to augment median star formation rates in those regions favorably relative to other parts of the planes, resulting in the behavior seen in figure~\ref{fig:ssfr}.  In any event, as these differences between our full and mass-limited samples remain small, all the important results we have discussed so far apply equally to both samples.

\section{DISCUSSION}

\subsection{Implications for Blue to Red Sequence Evolution}

The results discussed in the previous section point us towards a certain scenario of galaxy evolution.  The strong correspondence between color and morphology suggest that the two characteristics evolve concurrently as we, indeed, observe no red disk-dominated galaxies and few bulge-dominated blue ones.  The scenario is more ambiguous, however, when one looks at the star formation rate, absolute or specific, as a function of morphology.  This is due to the fact that, as we have shown, there exists a segregation in the star formation rate of bulge-like galaxies with, on one hand, dusty nuclear starbursts (blue compact galaxies can also have high star formation rates), and, on the other hand, old quiescent bulges.  This split is not present in color since, except for the blue compact galaxies, all of the above types of objects are red, which is why color shows a better correspondence with morphology than does specific star formation rate.  This implies that neither color nor morphology alone are good indicators of the physical state of a system.  Instead, geometry and dust conspire to link color and morphology in a way that appears independent of the star formation history.

The fact that many among the strongest star-forming galaxies possess bulge-like or near bulge-like morphologies, as measured by their {\em Gini} coefficient, indicates that episodes of strong star bursts are often linked to bulge growth.  \citet{Mihos96} showed that gas-rich major mergers, that can transform disk galaxies into ellipticals, trigger star bursts.  \citet{Robertson06} further showed that even when the gas is not entirely consumed and that a disk remains at the end of the process, considerable bulge growth still occurs.  It is thus likely that many of our strongly star-forming galaxies are late-stage mergers, and red-sequence objects in the becoming.  The fact that we also observe many of these high-SFR objects to be red is consistent with them often being IR-luminous \citep[e.g.][]{Sanders96}.  \citet{Chakrabarti06} indeed showed that gas-rich mergers emit most of their bolometric luminosity in the IR after the first pass due to infall of gas towards the center causing increased extinction.  Such merger events therefore appear to preserve the color-morphology correspondence mentioned above, and could explain our observations.

Semi-analytical models \citep{DeLucia06} indicate that some low to intermediate-mass field ellipticals ($M_{*} \le 10^{11} M_{\odot}$) could have built a non-negligible fraction of their mass from a last major merger event at redshifts as low as $z = 0.5$.  This is thus consistent with such objects being present in our sample.  On the observational front, \citet{Bundy05} found that, although the integrated stellar mass density evolved only mildly since $z \sim 1$, its morphological mix underwent significant changes, also hinting towards a significant occurrence of merging since that redshift.  Studies of B and R-band luminosity functions derived from the DEEP2 \citep{Faber05} and COMBO-17 \citep{Bell04,Wolf03} surveys similarly point towards such a scenario.  

Others, on the other hand, find evidence that the most massive galaxies were already in place by redshift $z \sim 1$ \citep{Juneau05,Treu05,Feulner05,Brinchmann00}.  We, on the contrary, observe that $\sim 30\%$ of our objects with mass $\log M_{*}/M_{\odot} > 10.5$ are experiencing significant star formation (SFR $> 1  \mbox{ M}_{\odot} \mbox{ yr}^{-1}$), in close agreement with \citet{Bell05}, and in support of the scenario in which a fraction of red sequence galaxies continue to grow after $z \sim 1$.

\subsection{Nature of Starbursting and Blue Compact Galaxies}

In order to investigate the nature of the starbursting galaxies in our sample, we pulled out all galaxies with a SFR $\ge 30 \mbox{ M}_{\odot} \mbox{ yr}^{-1}$, and performed a visual classification on them.  We find $\sim 30\%$ of them to be normal spirals, 25\% appear to be ongoing a major merger event, while another $\sim 35\%$ have bulge-like morphologies with either a clear underlying disk or features indicating recent events.  The remaining 10\% of our high SFR objects appear to possess either compact or regular elliptical morphologies, and could be AGN contaminations, or simply nuclear starbursts without any morphological peculiarities.

As mentioned in the previous section, strongly star-forming galaxies with bulge-like morphologies could be objects in late stages of a merger where only one newly formed galaxy remains with a strong nuclear starburst, a large bulge and most debris having either fallen back onto the galaxy or having been dispersed.  Alternatively, it is also possible that some of these bulges are being formed by other means such as secular instabilities \citep[e.g.][]{Debattista06,Wang06,Kormendy04}, also suggested by the fact that most of our star bursting late-type spirals possess bars.  Nonetheless, the existence of these objects appears robust as they have also been observed in infrared studies.  For example, \citet{Bell05} classify a fraction of their strongest starbursts into the E/S0 category.  \citet{Melbourne05} similarly finds a significant fraction ($\sim 20\%$) of luminous infrared galaxies at $0.61 < z < 1.00$ to possess compact or elliptical morphologies, and  \citet{Zheng04} finds $\sim 25\%$ of their LIRGs to have compact morphologies.  Although it is difficult to compare our UV-based measurements with these numbers, they are still mostly consistent.  The presence of such objects has implications on star formation quenching time-scales, and might pose some contradiction with the merger simulations of \citet{diMatteo05} in which the gas expulsion and star formation cutoff occurs before the two galaxies fully come together and settle into a spheroidal system.

We also mentioned the existence in our sample of a population of blue compact galaxies, a term we have loosely used to describe blue galaxies (FUV - $g \lesssim 2$) with either high concentration or {\em Gini}, or low asymmetry, clumpiness or M20.  Although this is a somewhat different definition from the one usually used in the literature \citep[e.g.][and references therein]{Noeske06} which is based, among other things, on size (typically $r_{50\%} \le 3.5$ kpc) and surface brightness (typically $\bar{\mu}(<r_{50\%}) \le 21 \mbox{ B mag arcsec}^{-2}$), most objects we characterize as blue compact galaxies do possess half-light radii and surface brightnesses that fit the above criteria.  These objects are thought to be dwarf galaxies brightened by intense star formation \citep[e.g.][]{Noeske06, Guzman03,Guzman97}.  A few blue concentrated galaxies, however, do show extended components, and could represent, as suggested by \citet{Hammer01} and \citet{Barton01}, bulges of spirals, in formation.  We cannot rule out, however, the possibility that some of them harbor weak AGNs which would confer them their blue FUV - $g$ color.  This question will be further investigated in future work.

\subsection{How Independent are Measurements of Star Formation Rate and Morphology?}

Throughout this paper we have attempted to look at how star formation rate and galaxy morphology correlate.  The question can be asked however as to whether these are truly independent quantities, or simply two measurements of the same phenomenon.  Is the light distribution representative of the structure of the galaxy?  Evidently, part of the answer is no, as star formation regions tend to stand out on top of the underlying structure.  This effect becomes all the more prominent as one moves to bluer bands.  In addition, significant structural components of a galaxy might be hidden under low surface brightness, and elude detection.

Clumpiness, for one thing, is precisely an attempt to pull out the star-forming regions from the underlying structure so as to obtain an estimate of the fraction of the light of the galaxy that is due to star formation, which is basically specific star formation rate.  Many factors (dust, inclination, resolution, nuclear star formation, etc.), however, contribute to create scatter in this simple relation, as described in the appendix, so that the correspondence between clumpiness and sSFR is essentially washed out.

Other morphological parameters are, in principle, also affected by the presence of star-forming regions. Asymmetry is probably the parameter that is the second most sensitive to star formation regions, and it is actually possible that the continuous rise of sSFR from A = 0.1 (where disks start to dominate) to A = 0.4 (where the relation becomes flat) (figure~\ref{fig:sfr-vs-casgm-complete-quantiles}) be accounted for by the presence of star formation regions.
On the other hand, concentration, {\em Gini} and M20, unlike S and A, are also sensitive to the presence and size of bulges.  The Gini coefficient, however, is affected by disk star-forming regions differently from how concentration and M20 are.  Compared to a smooth profile, adding bright regions far from the center of the galaxy would systematically decrease concentration, but it would, on the contrary, increase {\em Gini}, though star formation by itself can only increase {\em Gini} up to a certain point in the best of cases (which is about $G \sim 0.5$).  This behavior means that, unlike concentration that can be low even in the presence of a bulge due to peripheral bright star forming clumps or structures, the Gini coefficient systematically increases with the prominence of the bulge.  We infer that this is the reason for the correlation between the Gini coefficient and stellar mass, as bulges dominate at high masses and disks at low masses \citep{Kauffmann03b, Bundy05}.  The Gini coefficient is, thus, among our five parameters, the one that measures structural characteristics the most purely, as opposed to measuring star formation, or a combination thereof.  In that context, the correlation between star formation rate and the Gini coefficient in star-forming galaxies can be simply explained by the fact that star formation is, in most cases, more intense in nuclear starbursts and circumnuclear regions than it is in disks \citep[e.g.][]{Kennicutt98}.

Is the fact that different morphological parameters are differently affected by young stellar regions a problem for our investigation?  The answer depends on what we are trying to measure with these parameters.  We have shown in this paper that bulges and disks separate well in planes of asymmetry or clumpiness versus a concentration-like parameter (C, G and M20), and that that separation is fairly independent of the state of star formation in the galaxy as, for example, bulges segregate equally whether they are old or in the making through intense starbursts.  Similarly, the asymmetry parameter can be just as large for dry mergers as for gas-rich ones.  Non-parametric morphological parameters thus appear to represent the broad, overall structure of galaxies well enough.  Their usefulness diminishes, however, once one attempts to study more detailed morpholgy (e.g. spiral structure, interaction [other than major merger] signatures, etc.), as most parameters can be equally influenced by numerous different configurations, thus resulting in scatter in relations between morphological and physical parameters, as we have discovered throughout this work.

\section{SUMMARY}

In an age of ever expanding ground-based local (SDSS, 2dF), and spaced-based redshift (COSMOS, GEMS, GOODS) surveys,  it has become important to be able to understand and apply automatic galaxy classification.  In this paper, we have studied more specifically the concentration(C), asymmetry(A) and clumpiness(S) parameters as well as of the Gini coefficient(G) and the second moment of the brightest 20\% of the light(M20) of a sample of $z \sim 0.7$ galaxies in the COSMOS field.

We observe the galaxy population at redshift $z \sim 0.7$ to display a bimodality in both asymmetry and clumpiness, corresponding to a separation of the bulge-like and disk-like galaxies.  This partition is further accentuated when two morphological parameters are taken in combination, especially asymmetry or clumpiness with a concentration-like parameters (Concentration, Gini or M20).  Compared to local galaxies, we observe $z \sim 0.7$ galaxies to display a similar distribution of morphologies, although we see tentative evidence for them to be, on average, a little ($0-50\%$) more asymmetric.

By comparing with local samples of \citet{Kauffmann03b} and \citet{Brinchmann04}, we observe $z \sim 0.7$ galaxies to also have similar masses, sizes and surface mass densities, but higher star formation rates by a factor of 0.35 to 0.55 in log SFR.  As 0.55 is precisely the observed decrease in star formation rate density from $z \sim 0.7$ to present \citep{Schiminovich05}, we conclude that this overall decrease in the SFR of disk galaxies, or dimming, accounts for most of the star formation density evolution in the Universe in the last 6 Gyr.   We thus corroborate \citet{Wolf05} and \citet{Bell05}, who reached the same conclusion by respectively showing the $z \sim 0.7$ UV and IR luminosity functions to be dominated by normal spiral galaxies, implying that the present-day luminosity functions could be mostly reached by simple dimming of their studied population.  Our observed shift in SFR could be, however, on the low side (i.e. smaller than 0.55, the overall shift in star formation rate density), which suggests some degree of number evolution along with a decrease in the individual star formation rates.  Such a scenario is most consistent with the observed evolution of the UV luminosity function \citep{Arnouts05}, which, besides brightening, also includes a moderate steepening of the faint-end slope with redshift.

We also looked at relations between physical and morphological parameters and find them to possess much scatter.  We discussed the origin of this scatter for each of our morphological parameters, and discover that it is multi-faceted.  In particular, we find that star formation, in certain configurations, can have an effect on derived morphological parameters.  The resulting combination of star-forming regions and underlying structure that is measured can thus be, sometimes, hard to interpret.  We therefore conclude that the set of morphological parameters discussed in this paper, although very useful in describing the overall shape of galaxies as demonstrated, for example, in the bimodality, is not as well suited to study detailed morphology.

Nevertheless, we find {\em Gini} to correlate strongly with stellar mass which we understand to be a consequence of the fact that the Gini coefficient better traces the overall structure of a galaxy than any of the other morphological parameters we studied.  We also observe {\em Gini} to be the only morphological parameter with which the SFR of star-forming galaxies correlates.  This correlation further indicates that many of the strongest starbursts reach values of G similar to that of early-type galaxies, which otherwise usually have higher {\em Gini} than star-forming objects.  We see this tendency for starbursts (SFR $> 10 \mbox{ M}_{\odot} \mbox{ yr}^{-1}$) to carry bulge-like morphologies in our other parameters as well, and interpret those observations as evidence that episodes of strong star formation, proposed as a way to move galaxies from the blue to the red sequence \citep{Bell04,Faber05}, are often linked to the growth of a central component.

We find most of our starbursting galaxies with early-type morphologies to have red colors and large UV extinction coefficients, implying that they are substantially dust-enshrouded.  As a consequence, when looking at our full sample, we observe a strong correspondence between our morphological and FUV - $g$ color bimodalities.  Because red bulge-dominated galaxies are populated by both old and quiescent galaxies on one hand, and dusty and star-forming ones on the other, no such strong correspondence exists between morphology and specific star formation rate.  Color and morphology are, therefore, not necessarily good probes of the physical state of a galaxy, but their correlation implies that morphological and color evolution predominantly occurs concurrently.

The most plausible scenario we envisage to explain our results is one where bulges grow through episodes of intense, concentrated, gas and dust-rich episodes of star formation, whether from mergers or secular instabilities, that confer the galaxies both their red colors and bulge-like morphologies as well as their high star formation rates.  As intense starbursts trigger efficient feedback \citep{diMatteo05}, the star formation finds itself quenched after some time, and the galaxy quickly moves to the region of quiescent ellipticals.  During that process, both its morphology and color change little since, once star formation stops, the FUV-flux rapidly fades (in about 300 Myr) and the galaxy remains red.

Our interpretation implies that a fraction of the stars in the red sequence have formed at $z \lesssim 0.7$, and that the red sequence has kept growing in the last 6 Gyr through episodes of intense star formation, many of which are likely linked to merger events.

\acknowledgments

We are grateful to the anonymous referee for his/her thorough reading of the manuscript and detailed recommendations that significantly improved the paper.  Michel Zamojski would also like to thank Benjamin Johnson for stimulating and instructive discussions as well as Jarle Brinchmann for very helpful clarifications.

The HST COSMOS Treasury program was supported through NASA grant
HST-GO-09822. We wish to thank Tony Roman, Denise Taylor, and David 
 Soderblom for their assistance in planning and scheduling of the extensive COSMOS 
 observations.
 We gratefully acknowledge the contributions of the entire COSMOS colaboration
 consisting of more than 70 scientists. 
 More information on the COSMOS survey is available \\ at
  {\bf \url{http://www.astro.caltech.edu/$\sim$cosmos}}. It is a pleasure the 
 acknowledge the excellent services provided by the NASA IPAC/IRSA 
 staff (Anastasia Laity, Anastasia Alexov, Bruce Berriman and John Good) 
 in providing online archive and server capabilities for the COSMOS datasets.
 The COSMOS Science meeting in May 2005 was supported in part by 
 the NSF through grant OISE-0456439.
 
GALEX (Galaxy Evolution Explorer) is a NASA Small Explorer, launched in April 2003.
We gratefully acknowledge NASA's support for construction, operation,
and science analysis for the GALEX mission,
developed in cooperation with the Centre National d'Etudes Spatiales
of France and the Korean Ministry of
Science and Technology.

 {\it Facilities:} \facility{HST (ACS)}, \facility{GALEX}, \facility{Subaru}, \facility{CFHT}, \facility{KPNO}, \facility{CTIO}.

\bibliographystyle{apj}
\bibliography{apj-jour,ms}

\begin{appendix}
\section{Explaining the Noisy Relation Between Clumpiness and Specific Star Formation Rate}

Intuitively, one would think that clumpiness would be the parameter that best traces the specific star-formation rate.  However, as we have seen in figures~\ref{fig:sfr-vs-casgm} through~\ref{fig:sfr-vs-casgm-complete-quantiles} the relation has so much scatter that all correlation, except at the low-end (S $\lesssim 0.12$), is washed out.  This is confirmed in figure~\ref{fig:ssfr} by the histograms of sSFR vs S, as well as by the contours of constant specific star formation rate that run largely parallel to the S-axis in all plots.  To get a better insight into what clumpiness means physically, it is instructive to look at S vs A in figure~\ref{fig:color}, which shows that sometimes clumpier objects actually have redder colors than less clumpy ones.  As mentioned in section 4.1, high $S/A$ ratios indicate edge-on galaxies which, because of dust lanes and/or the use of a circular filter in the clumpiness algorithm, carry high values of S.  They are, however, often symmetric, and thus carry only moderate values of A.  Also, because of those same dust lanes, they emit less UV light in our direction than would a face-on galaxy.  On the other hand, some galaxies with blue cores carry diffuse asymmetric features.  These galaxies would typically have higher asymmetry as well as high UV-fluxes, but would not necessarily have high clumpiness values.  These could explain why sSFR values are often higher at low $S/A$ ratios and lower at high $S/A$ ratios.  They also contribute to reverse the expected correlation of sSFR and S.  Nevertheless, it is unlikely that inclination effects and diffuse asymmetric components are able to entirely account for the wide scatter in the sSFR - S relation.  We have seen in section 4.3 that clumpiness shows a suspicious correlation with size.  It is surprising because all of our morphological parameters are normalized quantities.  Such a correlation with size can therefore suggest resolution issues, which we now turn to to explore.

Although \citet{Elmegreen99} observed that a relatively constant fraction ($\approx 7\%$) of the B-band luminosity of late spirals and irregulars originates from star-forming complexes, we observe that spiral arms seem to often contribute as much to clumpiness as star-forming complexes themselves.  As density-wave structures are predominantly present in large disks, they contribute to create the observed trend.  Moreover, \citet{Elmegreen96} showed the size of the largest star-forming complexes, that is of the largest clumps, to be more or less proportional to the size of the galaxy, and \citet{Elmegreen99} further showed the number of complexes of a given diameter in a galaxy to be roughly proportional to the inverse square of their diameter, while their luminosity goes as the square of their diameter.  This implies two things.  First, we might not be able to resolve any clumpiness in some of our smallest galaxies, and secondly, even in galaxies for which we do resolve the largest complexes, since the product $n_{complex} \times L_{complex}$ is more or less independent of size, the total luminosity we measure in clumps roughly goes as $L_{D} \times \{\log D_{max} - \log D_{min}\}$, where $L_{D}$ represents a luminosity per dex of size and is multiplied by the difference of the log of the diameter of the largest complexes and that of the smallest ones we can resolve.  Since our resolution is constant, the total measured luminosity of star-forming complexes simply goes as $L_{D} \times \log D_{max}$, which in turns is proportional to $L_{galaxy} \times \log D_{galaxy}$, meaning that, since S is normalized to the total flux of the galaxy, $S \sim \log D_{galaxy}$, or $S \sim \log r_{50\%}$.   The correlation between clumpiness and size is thus, at least partly, the consequence of a lack of resolution.  This effect adds to the flattening of the sSFR vs. S relation as larger disks tend to have lower specific star formation rates (see figure~\ref{fig:rmuss}, or \citeauthor{Brinchmann04} \citeyear{Brinchmann04}), and is actually probably an even more significant source of scatter than inclination effects or morphological diversity.  Because clumpiness is influenced by these many factors, we question its usefulness as a star formation indicator.

\end{appendix}

\end{document}